\documentclass[twocolumn]{revtex4-2}
\usepackage{epsfig,graphicx}
\usepackage{amssymb,amsfonts,amsmath,bm}
\usepackage[colorlinks=true, linkcolor=blue, urlcolor=blue]{hyperref}

\usepackage{xr-hyper}
\begin{document}

\title{Nonparametric Reaction Coordinate Optimization with Histories for Rare Event Dynamics}
\author{Polina V. Banushkina$^1$, Sergei V. Krivov$^{1,2,*}$}
\affiliation{$^1$Faculty of Biological Sciences, University of Leeds, Leeds LS2 9JT, United Kingdom}
\affiliation{$^2$Astbury Center for Structural Molecular Biology, University of Leeds, Leeds LS2 9JT, United Kingdom}
\email{s.krivov@leeds.ac.uk}

\begin{abstract}
Rare but critical events in complex systems, such as protein folding, chemical reactions, disease progression, and extreme weather or climate phenomena, are governed by complex, high-dimensional, stochastic dynamics. Identifying an optimal reaction coordinate (RC) that accurately captures the progress of these dynamics is crucial for understanding and simulating such processes. However, determining an optimal RC for realistic systems is notoriously difficult, due to methodological challenges that limit the success of standard machine learning techniques. These challenges include the absence of ground truth, the lack of a loss function for general nonequilibrium dynamics, the difficulty of selecting expressive neural network architectures that avoid overfitting, the irregular and incomplete nature of many real world trajectories, limited sampling and the extreme data imbalance inherent in rare event problems. Here, we introduce a nonparametric RC optimization framework that incorporates trajectory histories and circumvents these challenges, enabling robust analysis of irregular or incomplete data without requiring extensive sampling. The power of the method is demonstrated through increasingly challenging analyses of protein folding dynamics, where it yields accurate committor estimates that pass stringent validation tests and produce high resolution free energy profiles. Its generality is further illustrated through applications to phase space dynamics, a conceptual ocean circulation model, and a longitudinal clinical dataset. These results demonstrate that rare event dynamics can be accurately characterized without extensive sampling of the configuration space, establishing a general, flexible, and robust framework for analyzing complex dynamical systems and longitudinal datasets.
\end{abstract}
\maketitle

Many rare yet important events are governed by complex, high-dimensional stochastic dynamics. Understanding and simulating such events can be greatly simplified by identifying a variable or index, often called a reaction coordinate (RC), that accurately captures the progress of the system's dynamics \cite{banushkina_optimal_2016, peters_reaction_2016}. In chemical kinetics, for instance, understanding rare events is important for such processes as chemical reactions, crystallization, and protein folding and aggregation. In these systems, knowing an appropriate RC enables the identification of transition states, the critical points or bottlenecks, that control the reaction dynamics. Moreover, an accurate RC can be used to accelerate or enhance the sampling of rare events in simulations \cite{jung2023machine, kang2024computing}.

For an RC to provide a faithful description of kinetics and to be practically useful for sampling, it should be optimally selected to minimize information loss resulting from dimensionality reduction. One prominent example of such an optimal RC is the committor function (also known as the splitting probability, or $p_\text{fold}$), which quantifies the likelihood that a given configuration will evolve towards a particular final state before reaching an alternative state. While originally developed in the context of chemical kinetics \cite{onsager_initial_1938, du_transition_1998, krivov_reaction_2013}, the concept of the committor has since been applied to problems across other fields. For example, it has been used to assess the probability of extreme weather or climate events \cite{lucente2022committor, jacques2023data}, to serve as an optimal biomarker to predict a patient's likelihood of recovery based on current conditions \cite{krivov_optimal_2014}, and to estimate the probability of winning a game of chess from a given position \cite{krivov_optimal_2011}.

Finding the committor for realistic systems is a notoriously difficult problem. The committor is a complex, non-linear function, formally defined as the solution of a high-dimensional partial differential equation over the system's configuration space.  Many promising approaches have been proposed to determine the committor, with recent advances driven by machine learning (ML) \cite{banushkina_nonparametric_2015, krivov_protein_2018, krivov_blind_2020, krivov_nonparametric_2021, li2019computing, fu2024collective, chen2023committor, khoo2019solving, aristoff2024fast, khoo2019solving, lucente2022committor, chen2023discovering,jung2023machine, kang2024computing, peters_obtaining_2006,best_reaction_2005, lucente2022coupling}. Despite the remarkable success of ML techniques in a variety of domains, for example, protein structure prediction \cite{jumper2021highly} or design \cite{watson2023novo}, progress in applying ML to learn the committor in complex systems has so far been relatively limited. Most existing studies have focused either on low-dimensional model systems or on realistic systems with relatively simple RCs and free energy landscapes featuring a single barrier. Reproducible analyses of systems with complex RCs and free energy landscapes involving multiple metastable states, which can serve as rigorous benchmarks to drive the field's development, remain scarce \cite{bhakat2022collective}.   This limited progress reflects the fact that determining the committor and analysing rare-event dynamics present a set of unique methodological challenges that make the straightforward application of standard ML techniques difficult. Next, we briefly highlight the main challenges that complicate committor estimation and explain how they fundamentally differentiate this problem from generic ML tasks. 
 
\textit{No ground truth}. For complex, realistic systems, the true values of the committor function are generally unknown, which makes it difficult to assess the accuracy of computed committors. In principle, one can evaluate the committor directly from its definition by initiating a large number of trajectories from a given configuration and measuring the fraction that reach state B before state A  \cite{du_transition_1998, banushkina_optimal_2016} . However, this brute-force approach is prohibitively expensive and can only be applied to a limited number of configurations, not to entire trajectories, and is infeasible in applications where generating new trajectories is impossible, such as analyses of longitudinal patient data. This limitation likely explains why only a small number of studies have attempted to quantify the accuracy of computed committors in complex, realistic systems.

\textit{No valid loss/objective function for general nonequilibrium dynamics}.  While there are valid loss functions for certain specific cases, where the committor is the exact minimizer, such as equilibrium dynamics \cite{krivov_reaction_2013, lorpaiboon2026exact}, Langevin dynamics with a constant diffusion coefficient \cite{kang2024computing, rotskoff2021active}, or nonequilibrium settings in which every trajectory reaches a boundary state \cite{peters_obtaining_2006}, no such loss function is known for the general case of non-equilibrium dynamics, e.g., ensembles of short trajectories, often encountered in practical applications. As a result, standard ML practices such as train/test splits or cross-validation used to evaluate model performance or detect overfitting cannot be directly applied.

\textit{NN architecture}. Selecting an appropriate neural network architecture is challenging: on the one hand, the network must be sufficiently expressive to approximate the highly complex, high-dimensional committor; on the other hand, it must contain a manageable number of parameters to reduce the risk of overfitting \cite{lorpaiboon2026exact}. This difficulty is compounded by the fact that it is hard to assess approximation accuracy or reliably detect overfitting in the absence of a true loss function or ground truth. 

\textit{Irregular data}. Empirical trajectories, in contrast to those generated in controlled computer simulations, are usually highly irregular. For example, trajectories may be interval censored or right censored when patients miss appointments or drop out of a study. They often contain missing entries, inconsistent sampling intervals, or corrupted measurements.  Weather datasets exhibit similar challenges: observations are collected from heterogeneous sources with sparse, uneven sampling \cite{nguyen2023climax}. Single-molecule experiments may provide only very low-dimensional projections of the underlying dynamics \cite{kodera2021structural, bustamante2021optical, nettels2024single}. Standard ML methods are generally sensitive to such imperfections in the input data, making accurate committor estimation considerably more challenging in these settings.

\textit{Rare events}. 
Rare events dynamics datasets are, by definition, strongly imbalanced; the configurations associated with the events of interest, such as folding-unfolding transitions in protein folding trajectories, constitute only a tiny fraction of the full dataset. Consequently, performance must be evaluated using metrics that are sensitive to the accuracy of the committor specifically in these rare regions, since their contribution to aggregate metrics (e.g., mean squared error) may be negligible. For example, if only 1\% of configurations are misclassified, this may appear acceptable; however, for realistic ratios, e.g., when the population of the transition state (TS) is 1/1000 that of the native basin, this would imply that ten out of eleven putative TS configurations are actually native basin states misclassified as TS.

A second challenge is achieving convergence during batch based optimization, a standard ML technique \cite{zeng2021note, lorpaiboon2026exact}. Most batches will contain no or at most a few rare events of interest, making the gradient estimate uninformative.

Finally, in ML one typically requires that the model (the learned committor) generalize to unseen data. However, generalizability requires that the train and test data follow the same underlying distribution, which in turn requires extensive and representative sampling. Ensuring such coverage is particularly difficult for rare events in high-dimensional configuration spaces \cite{gabrie2022adaptive}. 

Progress in this research area requires addressing all of these challenges. We propose the framework of nonparametric RC optimization with histories as one possible way to do so. Specifically, we present methods for the robust and accurate determination of the committor function (and other optimal RCs) from irregular and incomplete longitudinal datasets. The non-parametric nature of the framework allows the committor to be approximated arbitrarily closely, and using the entire trajectory at each iteration avoids the small batch issues for rare events. A key ingredient of the framework is the set of highly sensitive validation criteria that assesses the accuracy of the putative committor (and other optimal RCs). Importantly, the criteria are not based on a single aggregate quantity: they verify that the putative RC satisfies the corresponding equation for every value of the RC and across all relevant time scales. Because the validation does not rely on train/test splits, it does not assume extensive or representative sampling of the configuration space. This, in turn, makes it possible to separate two tasks: (i) accurately determining the committor within the sampled region, and (ii) learning a global committor that generalizes to unseen configurations or the entire configuration space. The former is sufficient for simulation analysis and enhanced sampling \cite{krivov2023nonequilibrium}, whereas the latter would require exponentially more data and is typically infeasible.

 The power of the framework is thoroughly evaluated through increasingly challenging analyses of realistic protein folding dynamics, where it yields accurate committor estimates that pass the stringent validation criterion and produce high-resolution free energy profiles. Its generality is further illustrated across diverse systems, including phase space dynamics, conceptual ocean circulation models, and a clinical longitudinal dataset.

\section*{Results}
\subsection*{Nonparametric RC optimization with histories}
To introduce the idea of RC optimization, consider stocshastic dynamics in a multidimensional configuration space $\vec{X}$, describing a reaction between two specified boundary states, $A$ and $B$. Introduce an RC, $r(\vec{X}): \vec{X} \rightarrow [0,1]$, which projects $\vec{X}$ onto a one-dimensional coordinate while satisfying boundary conditions $r(\vec{X}\in A)=0$ and $r(\vec{X}\in B)=1$. Dynamics projected onto an RC are generally non-Markovian due to the loss of information associated with the dimensionality reduction. This non-Markovian behavior can be accurately described using the generalized Langevin equation (GLE) with a memory kernel, which can be derived via the Mori-Zwanzig formalism \cite{mori_transport_1965, zwanzig2001nonequilibrium}. In principle, the GLE can be derived for any reasonable coordinate; however, computing the memory kernel in practice is very difficult \cite{darve_computing_2009, klippenstein2021introducing}. 

Neglecting the memory term yields a simplified diffusive model, fully defined by the free energy $F(r)$ and diffusion coefficient $D(r)$ as functions of the RC. These quantities can be estimated from an equilibrium trajectory at the timescale of the trajectory saving interval \cite{banushkina_optimal_2016, krivov_protein_2018}. However, such models generally provide a poor description of system dynamics. They tend to have less complex free energy landscapes with lower apparent free energy barriers and predict faster kinetics, manifesting as a shorter mean first passage time (MFPT) or larger flux between the states $A$ and $B$ \cite{krivov_hidden_2004, krivov_diffusive_2008, banushkina_optimal_2016}.

To improve the accuracy of a diffusive model, the RC can be optimized to minimize non-Markovian effects. This involves perturbing the RC, $r'(\vec{X}) = r(\vec{X}) + \delta r(\vec{X})$, where $\delta r(\vec{X})$ vanishes at the boundary states $A$ and $B$ to preserve boundary conditions. If the modified RC yields improved kinetics, typically indicated by reduced flux, it is accepted. The optimization process is repeated until no further improvement is possible. 

The key question is how much such optimization can improve description of kinetics by mitigating memory effects. For equilibrium stochastic dynamics between two boundary states, the optimal RC is the committor function $q(\vec{X})$, which gives the probability that a system starting at configuration $\vec{X}$ will reach state $B$ before $A$. Remarkably, a simple diffusive model along the committor can \textit{exactly} compute key kinetic quantities such as flux, MFPT, and mean transition path time (MTPT) between any two points (iso-surfaces) on the committor \cite{krivov_protein_2018}. This result holds for arbitrarily complex equilibrium stochastic dynamics in configuration space and does not require a separation of timescales \cite{krivov_reaction_2013, lu_exact_2014, berezhkovskii_diffusion_2013}. In theory, iterative RC optimization should converge to the committor, yielding an accurate description of the system's kinetics.  

In practice, the optimization proceeds as follows. Given an RC $r(\vec{X})$ and a long equilibrium trajectory $\vec{X}(t)$, one can compute the RC time-series $r(t)=r(\vec{X}(t))$; where $t$, for brevity, denotes time-series sampled at regular intervals $k\Delta t$. From this, the free energy and the diffusion coefficient can be estimated, fully defining the diffusive model. Since the RC time-series alone determines the model, optimization can be performed directly on $r(t)$ rather than on the functional form $r(\vec{X})$. This simplification underlies the nonparametric approach to RC optimization \cite{banushkina_nonparametric_2015, krivov_protein_2018, krivov_blind_2020, krivov_nonparametric_2021}. Starting with the seed RC time-series, satisfying the boundary conditions, variations of the RC time-series are introduced as $r'(t)=r(t)+\delta r(t)$, where $\delta r(t)=f(r(t),y(t))$. Here, $f$ is a function of the RC itself and a randomly chosen collective variable (CV) time-series $y(t)=y(\vec{X}(t))$, and $\delta r(t)$ equals zero at the boundaries. The variation $f$ is selected such that it provides the greatest improvement to the putative RC time-series, for example, the one that minimizes the functional $\Delta r^2=\sum_t [r'(t+\Delta t) -r'(t)]^2$ \cite{krivov_reaction_2013,krivov_protein_2018}. If the optimization converges to the optimal RC we say the set of CVs is complete. For the committor in equilibrium systems, this happens when $\Delta r^2$ reaches its theoretical lower bound of $2N_{AB}$, where $N_{AB}$ is the number of transitions between boundary states in the long equilibrium trajectory \cite{krivov_reaction_2013}.

In practice, the available set of CVs is often incomplete. We show that  incorporating history, i.e., past trajectory segments, can help compensate for missing variables and improve the accuracy of the estimated RC. Specifically, we introduce variations of the form $\delta r(t) = f(r(t-\Delta t_h), y(t-\Delta t_h))$, where $\Delta t_h > 0$ is a time delay (see Methods for details). The idea is that past dynamics can reveal hidden patterns, such as distinguishing between parallel pathways, even when key variables are unobserved. This approach is conceptually related to Takens' embedding theorem, which shows that time-delayed coordinates can reconstruct the dynamics of complex systems from partial observations \cite{takens1981lecture}. 
A key advantage of the nonparametric framework is its simplicity: incorporating histories only requires expanding the set of variations. In contrast, parametric methods must explicitly model time-lagged  dependencies, for example, using an advanced neural network architecture.

\subsection*{Application to protein folding trajectory}
To demonstrate the power of the approach we apply it to protein folding dynamics, which arguably exhibit the most complex RCs and corresponding free energy landscapes. Specifically, we analyze a long equilibrium atomistic trajectory of HP35 protein \cite{piana_protein_2012} (see Methods for details). This system is an excellent benchmark, challenging yet tractable, with reproducible results \cite{krivov_protein_2018, krivov_blind_2020, nagel2023toward, sasmal2025improved, kim2024scalable}. It presents several characteristic difficulties typical of real-world problems. The configuration space is extremely high-dimensional ($\sim 3\times 577$ degrees of freedom). The events of interest, 146 folding-unfolding transitions, are rare, making their analysis and simulation particularly challenging. Limited sampling makes the transition state region especially prone to overfitting. Moreover, the large entropic contribution to the free energy, a characteristic feature of biomolecules, means exponentially many folding pathways within the folding funnel \cite{onuchic_protein_1996}. Consequently, it is unrealistic to expect extensive sampling, most pathways are typically sampled only a few times at best. 

We performed repeated analyses of the system, progressively increasing the level of difficulty to rigorously benchmark the performance of our approach in challenging scenarios. We began with an extensive, complete set of CVs, then used a smaller, incomplete set. Next, we analyzed a highly irregular ensemble of short trajectories generated by resampling the original trajectory. Finally, we tested the method using just a single CV - the RMSD time-series. Across all scenarios, the results were consistent with each other and with previous studies demonstrating the robustness of our approach. Technical details of the analyses are provided in the Supplementary Information.

\subsubsection*{A complete set of CVs}
\begin{figure}[htbp]
	\centering
	\includegraphics[width=\linewidth]{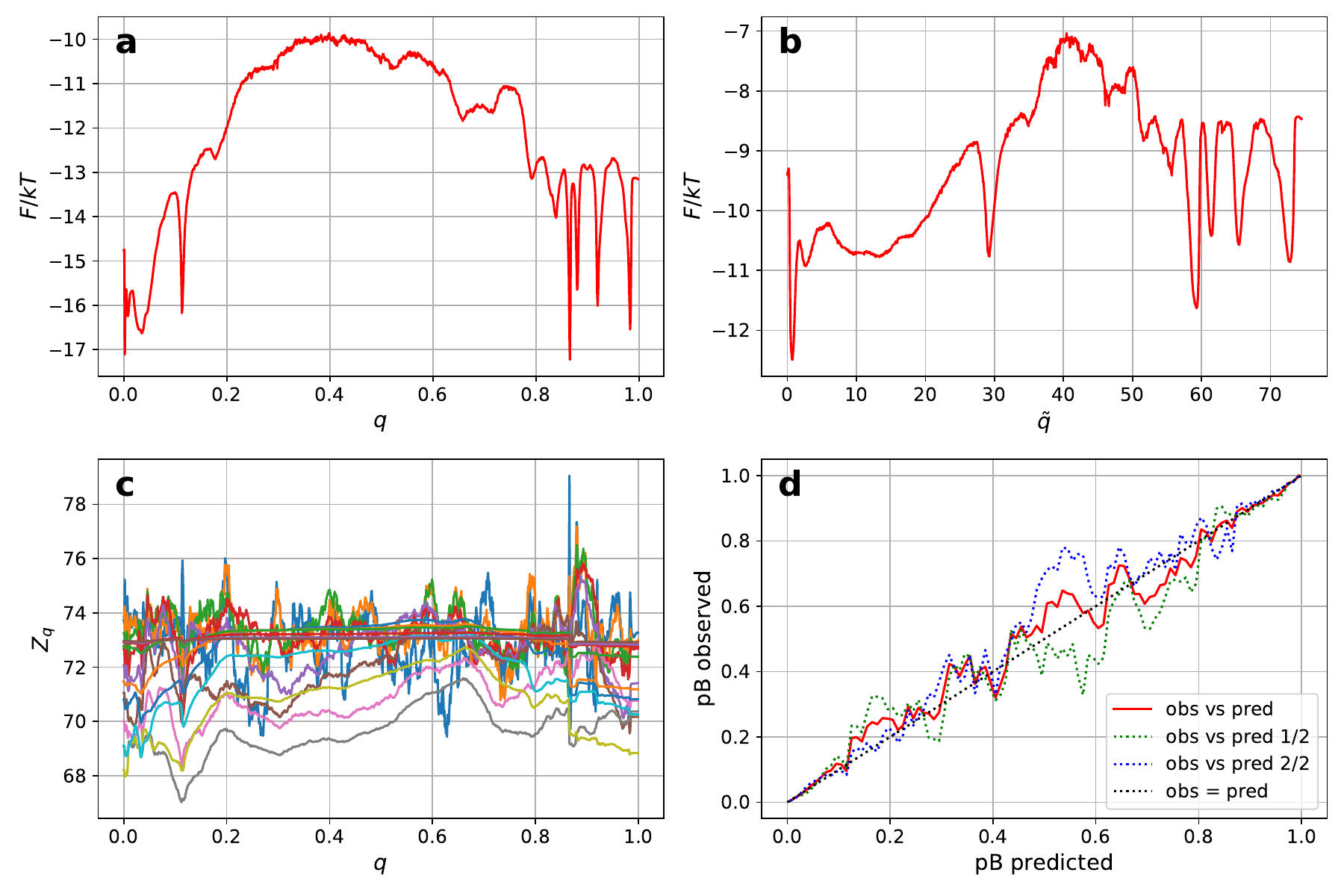}
	\caption{RC optimization with histories using a complete set of CVs. Free energy as a function of the committor $q$ (\textbf{a}) and the natural committor $\tilde{q}$, where $D(\tilde{q})=\text{const}$ (\textbf{b}). (\textbf{c}) Validation criterion $Z_q(q, k \Delta t_0)$ remains approximately constant across different lag times (time-scales) $k=1,2,...,2^{16}$, indicating  that the putative optimal RC closely approximates the committor. (\textbf{d}) Predicted vs. observed committors for the full trajectory, as well as for its first (1/2) and second (2/2) halves, show good agreement; the halves are used to estimate statistical fluctuations.}
	\label{fig:1}
\end{figure}

We begin with a complete set of CVs, allowing the nonparametric equilibrium optimization without histories to reach the theoretical lower bound of $\Delta r^2\sim 2N_{AB}$ (Supplementary Fig. \ref{fig:S1}). However the optimization was not uniform and led to overfitting in the TS region (Supplementary Fig. \ref{fig:S1}c). One way to mitigate this is through adaptive optimization, which focuses on less optimized regions \cite{krivov_protein_2018, krivov_blind_2020}. Here we show that incorporating histories makes the optimization more uniform and eliminates overfitting (Fig. \ref{fig:1}).   

Figs. \ref{fig:1}a and  \ref{fig:1}b show complex, high-resolution free energy profiles (FEPs) with multiple minima in both the denatured and native basins in excellent agreement with previous results obtained using different CVs and optimization methods \cite{krivov_protein_2018, krivov_blind_2020}. While the primary focus of this work is the accurate determination of the committor, a detailed interpretation of the FEPs in the context of folding dynamics lies beyond the scope of this study and has been thoroughly addressed in earlier publications.

The validation criterion $Z_q$ (see Methods) remains relatively constant (Fig. \ref{fig:1}c; cf. Supplementary Fig. \ref{fig:S1}c), within expected statistical fluctuations of approximately $10\%$, estimated as $1/\sqrt{2N_{AB}}$. Predicted and observed committors are in good agreement (Fig. \ref{fig:1}d; cf. Supplementary Fig. \ref{fig:S1}d), up to expected statistical fluctuations estimated by considering first and second halves of the trajectory. Importantly, $Z_q$ criterion exhibits a much larger difference between the optimal and sub-optimal RCs, compared with predicted vs. observed committor plots, highlighting its sensitivity and rigor. Additionally, the final $\Delta r^2 \approx 146$, matches the theoretical lower bound, confirming the internal consistency of the approach. Together these results demonstrate that the optimized RC closely approximates the committor.

The constancy of $Z_q$ across different lag times indicates that the diffusion coefficient, as a function of the RC, remains invariant when determined from the equilibrium trajectory at different lag times, a characteristic feature of Markovian description. In contrast, a sub-optimal RC yields a time-scale-dependent diffusion coefficient reflecting sub-diffusive, non-Markovian behavior \cite{krivov_is_2010}.

\subsubsection*{An incomplete set of CVs.}
\begin{figure}[h!]
	\centering
	\includegraphics[width=\linewidth]{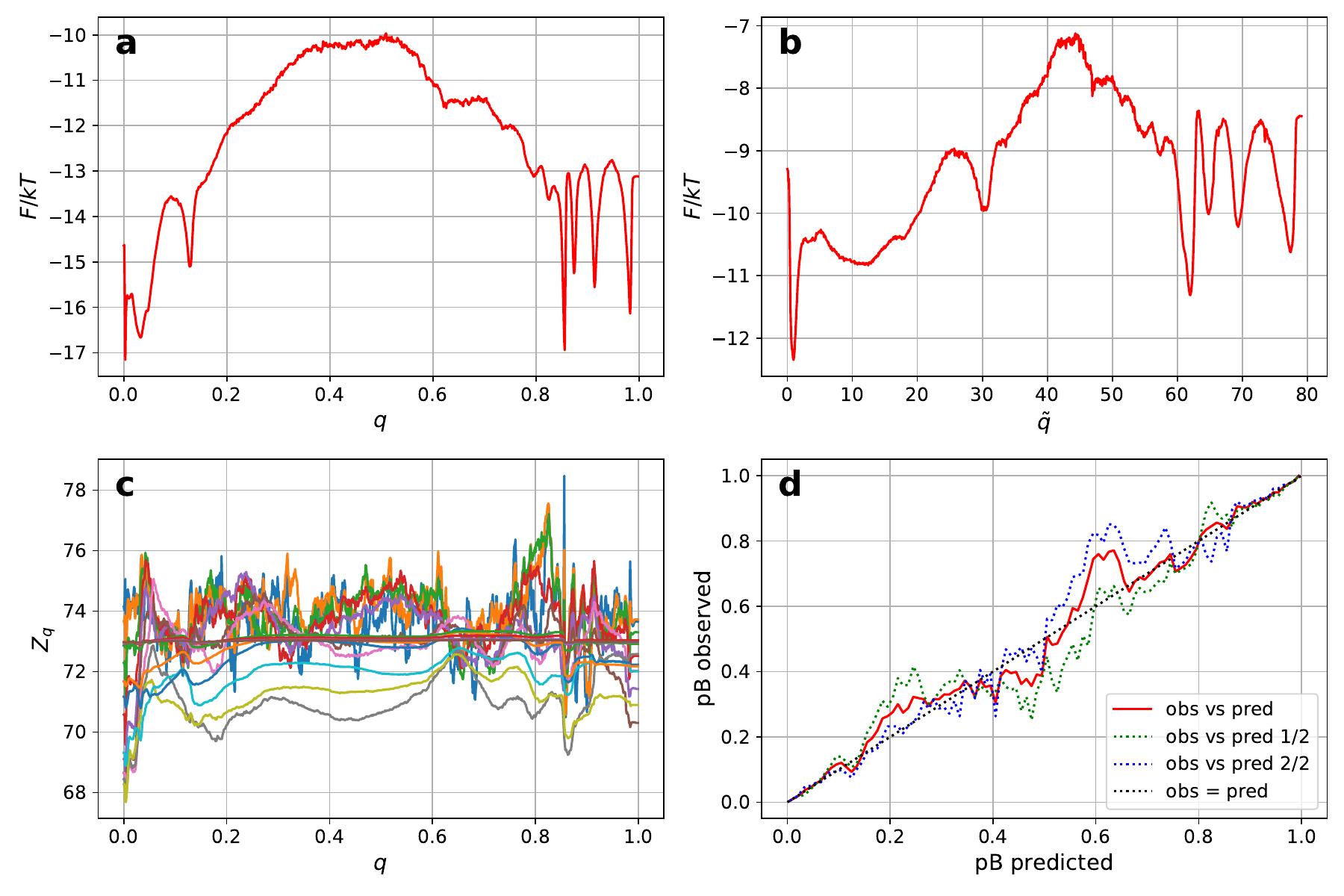}
	\caption{RC optimization using an incomplete set of CVs. Results are in very good agreement with those shown in Fig. \ref{fig:1}. Notation as in Fig. \ref{fig:1}.}
	\label{fig:2}
\end{figure}

We next consider an incomplete set of CVs, where nonparametric optimization without histories fails to reach the theoretical lower bound for $\Delta r^2$ (Supplementary Fig. \ref{fig:S2}). Such scenarios with missing information are common in practice. For example, in clinical datasets, not every variable can be measured, and simulations may omit solvent degrees of freedom to save disk space. We show that incorporating histories allows the optimization to compensate for missing information (Fig. \ref{fig:2}).

Figs. \ref{fig:2}a and \ref{fig:2}b show that the resulting FEPs closely match those in Fig \ref{fig:1}. The validation criterion (Fig. \ref{fig:2}c) and the predicted vs observed committor plot (Fig. \ref{fig:2}d) confirm that the computed RC closely approximates the committor.

\subsubsection*{A highly irregular longitudinal dataset.}
Here, we analyze a highly irregular ensemble of short trajectories, generated by resampling the original trajectory (see Methods). These trajectories vary in length, are saved with variable time intervals, and contain randomly distributed missing values, characteristics typical of clinical datasets. Similar challenges arise in weather and climate data, where observations are collected at different, relatively sparse spatial locations and at different temporal frequencies \cite{nguyen2023climax}. Analyzing such datasets is challenging for most machine learning approaches, which typically expect regular input. This ensemble can also be interpreted as arising from enhanced sampling techniques \cite{krivov_nonparametric_2021, jung2023machine, kang2024computing, kohlhoff_cloud_2014, lohr_abeta_2021}.

\begin{figure}[h!]
	\centering
	\includegraphics[width=\linewidth]{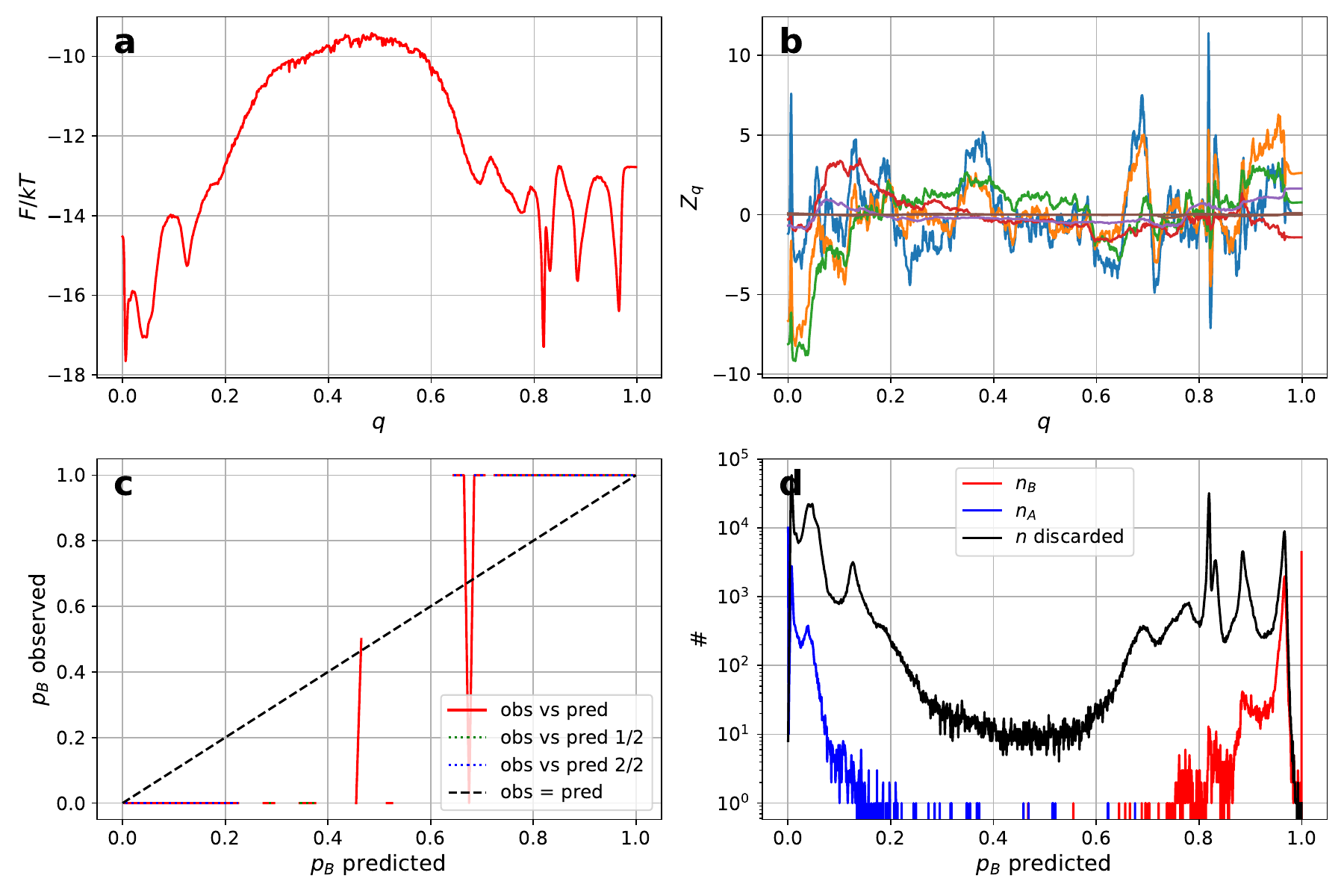}
	\caption{RC optimization for highly irregular ensemble of trajectories, mimicking a clinical dataset. Results are in good agreement with that shown on Figs. \ref{fig:1} and \ref{fig:2}. (\textbf{a}) Free energy as a function of the committor; (\textbf{b}) Validation criterion $Z_q$ is relatively constant up to statistical fluctuations; curves are shifted to have zero mean for clarity. (\textbf{c}) Predicted vs. observed committors disagree, as most of trajectories do not reach boundary states. (\textbf{d}) Histogram of trajectories reaching state $A$ ($n_A$), state $B$ ($n_B$), or neither ($n$ discarded).}
	\label{fig:3}
\end{figure}

Fig \ref{fig:3}a shows that the nonparametric approach with histories successfully recovers the committor and FEP closely matching those obtained from the original trajectory. Minor differences can be attributed to statistical fluctuations introduced during resampling. The validation criterion (Fig. \ref{fig:3}b) further confirms the optimality of the committor determined from the irregular dataset. 

In contrast, the predicted vs observed plot (Fig. \ref{fig:3}c) shows clear disagreement. This is because most trajectories in the ensemble do not reach either boundary state (Fig. \ref{fig:3}d), making standard evaluation metrics, such as the predicted vs observed plots or the area under the curve (AUC), biased and unreliable \cite{banushkina2025data}. The validation criterion $Z_q$, provides a more robust and consistent measure of RC optimality under these conditions.

\subsubsection*{A single CV}
Finally, we consider the extreme case of using only a single input variable, the RMSD time-series. Such low-dimensional inputs are common in applications. For example, in single-molecule experiments probing protein folding dynamics, accessible observables include force-extension profiles in the atomic force microscopy \cite{kodera2021structural}, the inter-dye distances in the F\"orster resonance energy transfer (FRET) measurements \cite{nettels2024single} and bead-to-bead separations in optical tweezers assays \cite{bustamante2021optical}. Similarly, wearable devices, such as smart watches and rings continuously provide low dimensional input streams related to patient health \cite{bayoumy2021smart}.

To further demonstrate the generality of the approach, we use the mean first passage time (MFPT) to the native state as an optimal RC. The optimization algorithm is modified by using the corresponding optimization functional \cite{krivov_nonparametric_2021}.
\begin{figure}[h!]
	\centering
	\includegraphics[width=\linewidth]{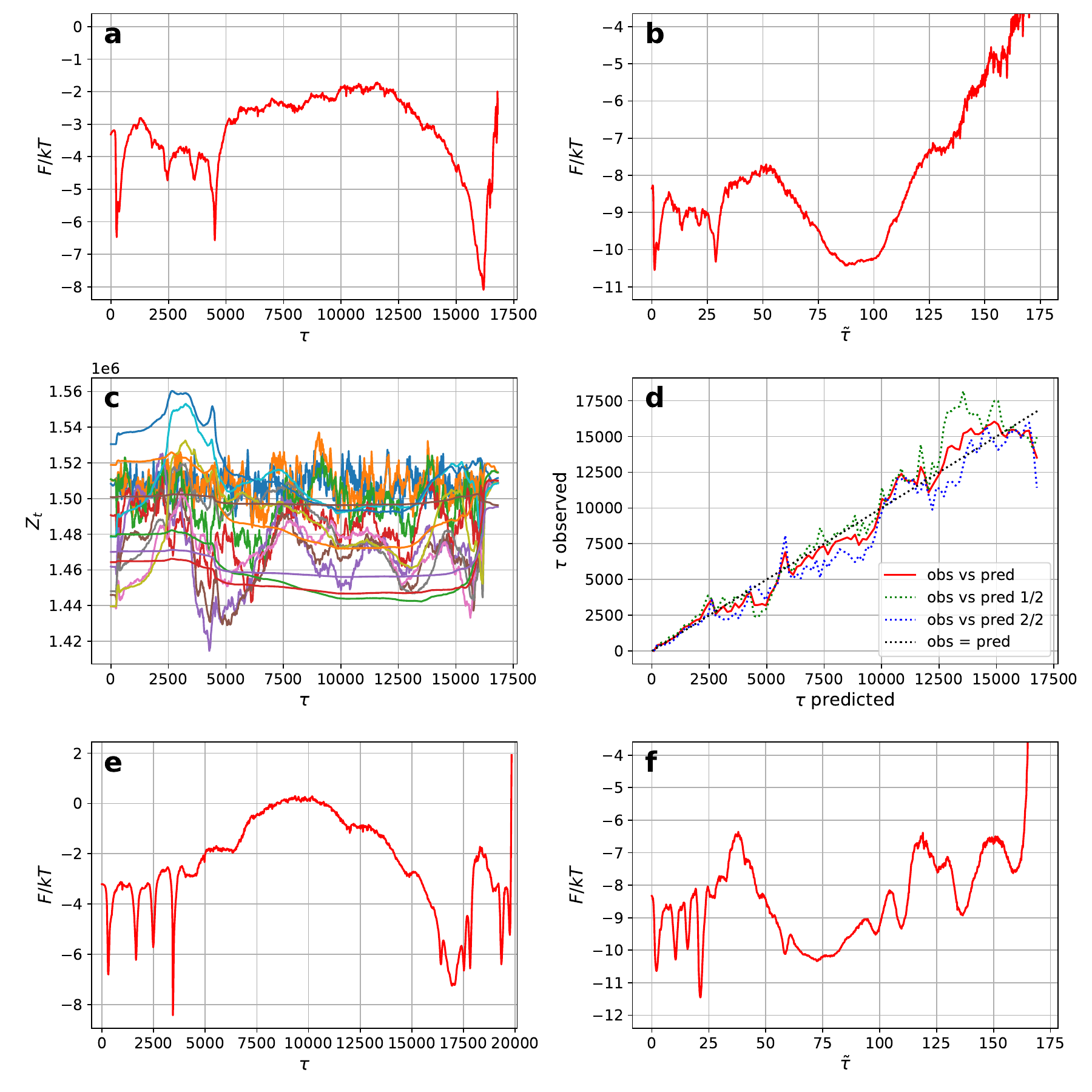}
	\caption{The MFPT RC optimization using a single variable - the RMSD time-series: Free energy as a function of the MFPT $\tau$ (\textbf{a}) and  the natural MFPT $\tilde{\tau}$, where $D(\tilde{\tau})=\text{const}$ (\textbf{b}). Note that native basin lies on the left, as the native state corresponds to $\tau=0$. (\textbf{c}) Validation criterion $Z_\tau$ is relatively constant ($\pm 4\%$), indicating that the putative optimal RC closely approximates the MFPT. (\textbf{d}) Predicted and observed MFPT are in good agreement. Free energy profiles along the MFPT (\textbf{e}) and natural MFPT (\textbf{f}), determined, using a complete set of CVs, for comparison.}
	\label{fig:4}
\end{figure}  

 Figs. \ref{fig:4}a and \ref{fig:4}b show the FEP as functions of MFPT and MFPT rescaled to the natural coordinate. The validation criterion (Fig. \ref{fig:4}c) is relatively constant, with slight deviation in the region $0<\tau<5000$, indicating mild sub-optimality. Predicted and observed MFPTs are in good agreement (Fig. \ref{fig:4}d). For comparison, Figs. \ref{fig:4}e and \ref{fig:4}f show FEPs for the MFPT RC obtained using the complete set of CVs. These are in relatively good agreement with Figs. \ref{fig:4}a and \ref{fig:4}b, considering the extremely low-dimensional input of the latter. Both FEPs capture similar large time-scale features, for example, the denatured minimum is around $\tau\sim 16000$, with difference mainly in the fine structure of fast dynamics. In contrast, optimization without histories yields significantly worse results (Supplementary Fig. \ref{fig:S4}); with MFPTs and time-scales off by orders of magnitude.

Fig. \ref{fig:4}f illustrates the benefits of transforming to the natural coordinate (cf. Fig. \ref{fig:4}e). Generally, the optimal RC exponentially enlarges the barriers and compresses the minima. The natural coordinate restores the relative sizes of the minima and barriers.

Fig. \ref{fig:4}f highlights an advantage of the MFPT RC over the committor: it requires only a single boundary state. Defining boundary states is not trivial. Typically, it involves using an order parameter, such as RMSD, which can obscure the inherent complexity of the free energy landscape of the boundary states \cite{krivov_blind_2020}. Using MFPT RC enables boundary states to be defined based on dynamics, rather than on an arbitrary chosen, albeit physically motivated, order parameters. This allows it to reveal many more minima in the denatured basins (cf. Fig. \ref{fig:1}b), reflecting the system's kinetics more accurately.

The structure of the FEP in the denature basin, found using the MFPT RC is in very good agreement with that identified by blind analysis using eigenvectors \cite{krivov_blind_2020}. The corresponding parts of FEPs along the committor (Fig. \ref{fig:1}b), the MFPT (Fig. \ref{fig:4}f) and the first eigenvector (Fig. 5a in \cite{krivov_blind_2020}) are in good agreement, indicating the robustness of the framework and the obtained results.  

Following rigorous testing, we now apply the method to diverse systems to demonstrate its broad applicability.

\subsection*{Committors for dynamics in phase space}
Until now, we have considered RCs as functions of configuration space. However, it might be useful to consider RCs as functions of phase space (positions and momenta), to analyze simulations at time-scales where the system retains memory of its velocities. In such cases, the dynamics are Markovian in the phase space rather than in configuration space. 

For example, one general strategy to overcome sampling problems in biomolecular simulations involves simulating a very large ensemble of short trajectories instead of a single long trajectory \cite{krivov_nonparametric_2021,jung2023machine,kang2024computing,kohlhoff_cloud_2014,lohr_abeta_2021}. Analyzing such trajectories at very short lag times can allow significantly shorter trajectories and higher efficiency. Here we show that RC optimization with histories, without explicitly considering the velocities (i.e., using configuration space as input), allows the determination of optimal RCs as functions of phase space. 
\begin{figure}[h!]
	\centering
	\includegraphics[width=\linewidth]{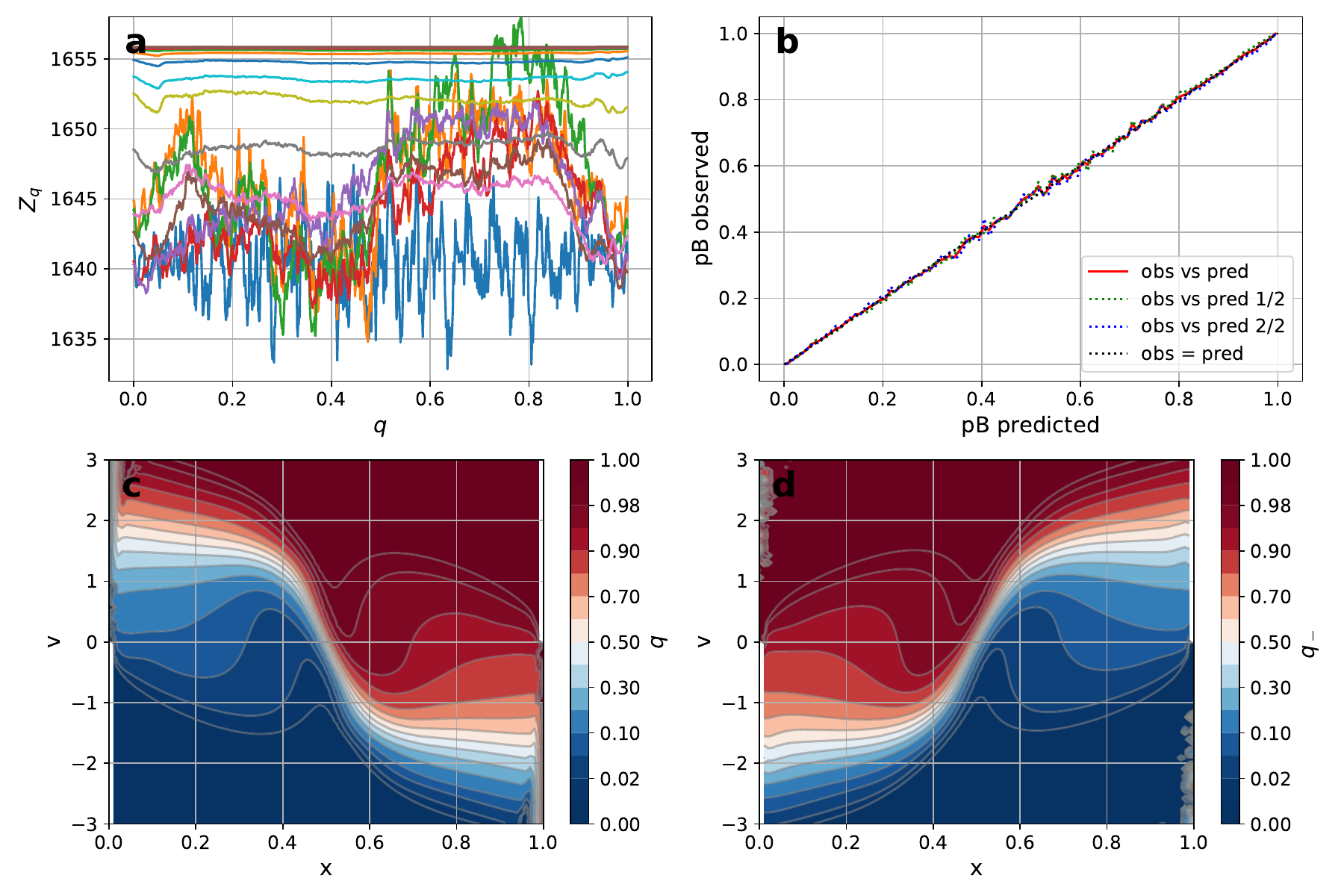}
	\caption{RCs as functions of phase space for underdamped Langevin dynamics: (\textbf{a}) Validation criterion ($Z_q$) remains approximately constant ($\pm 0.6\%$), up to statistical fluctuations. (\textbf{b}) Predicted and observed committors agree. (\textbf{c}) Committor as a function of phase space. (\textbf{d}) Backward committor $q_-$ as a function of phase space.}
	\label{fig:5}
\end{figure}
To illustrate this consider underdamped Langevin dynamics on the interval $0<x<1$ with reflecting boundaries at both ends and a barrier located at the center.

Fig. \ref{fig:5}a shows the validation criterion, confirming that the obtained RC closely approximates the committor. Predicted and observed committors are in excellent agreement (Fig. \ref{fig:5}b). 
Fig. \ref{fig:5}c reveals a nontrivial dependence of the committor on velocity (used here only for  visualization). 

In systems lacking detailed balance, both forward and backward committors are required to describe reactive trajectories \cite{e_towards_2006}. The latter, denoted by $q_-(\vec{X})$ is the probability that trajectory at point $\vec{X}$ came from state $A$ rather than $B$. Its time-series $q_-(t)$ can be determined by simply applying the nonparametric approach to the time-reversed trajectory. Fig. \ref{fig:5}d shows the resulting $q_-(x,v)$, which in this case, satisfies $q_-(x,v)=1-q(x,-v)$.  

\subsection*{Committor for a conceptual ocean model}
\begin{figure}[h!]
	\centering
	\includegraphics[width=\linewidth]{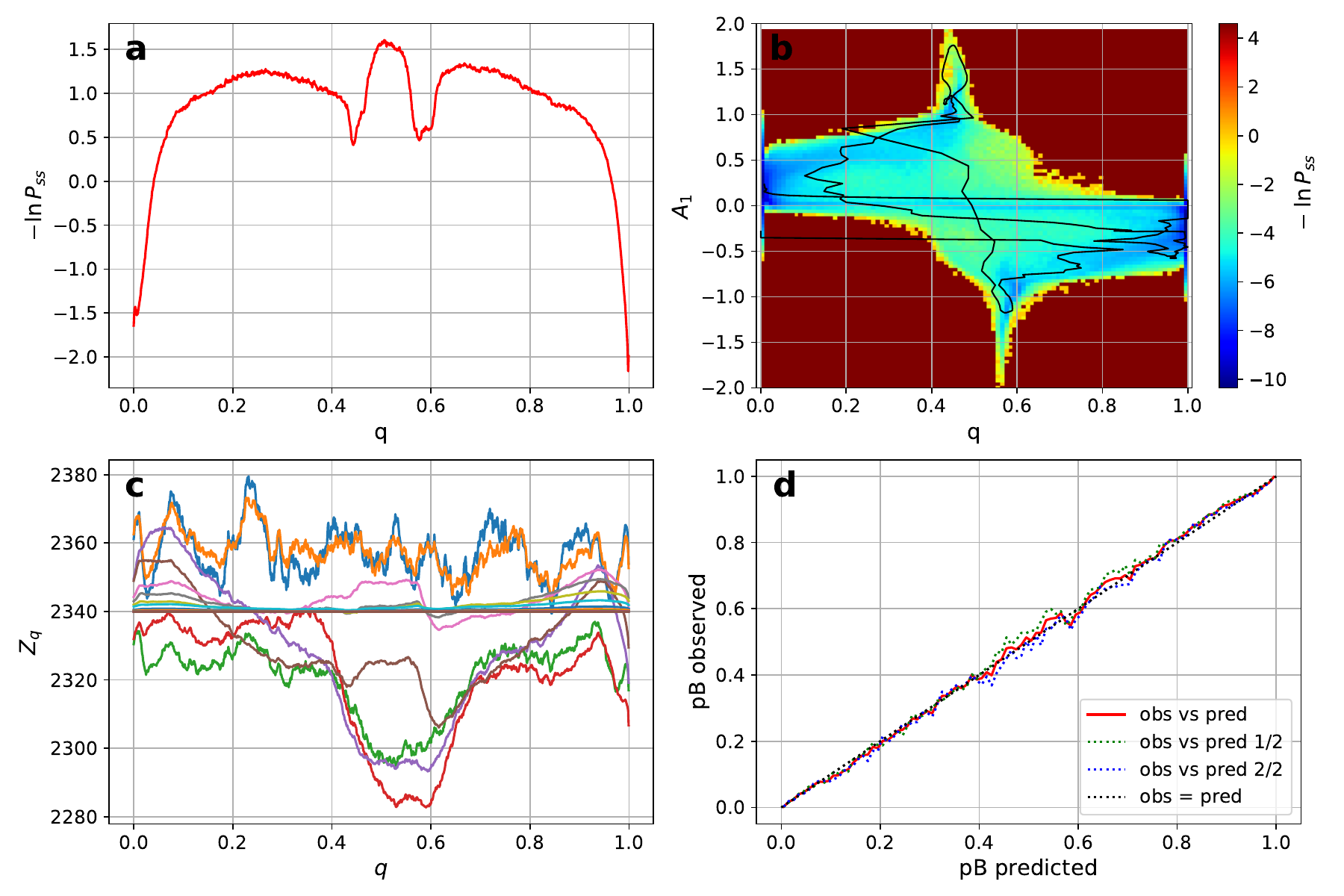}
	\caption{Committor optimization for the double-gyre model of the AMOC. (\textbf{a}) Steady-state probability as a function of the committor $q$ and (\textbf{b}) that as a function of $q$ and the model variable $A_1$, both revealing two marginally stable intermediate states. A segment of trajectory (black line) illustrates the complex dynamics. (\textbf{c}) validation criterion ($Z_q$) remains approximately constant ($\pm 1\%$), up to statistical fluctuations. (\textbf{d}) predicted and observed committors agree.}
	\label{fig:6}
\end{figure}

The Atlantic Meridional Overturning Circulation (AMOC) is a key component of the global climate system, responsible for large-scale heat and salt transport in the ocean. Conceptual models and observations suggest that the AMOC may exist in a bistable regime \cite{garzoli2013south}, making it susceptible to abrupt collapse under changing freshwater forcing \cite{rahmstorf1996freshwater}. Although such a collapse is considered a rare event, its potentially severe impact makes it critical to accurately analyze and simulate the underlying dynamics, which can be greatly assisted by knowing the committor function. Recent work has evaluated methods for computing the committor on two conceptual AMOC models \cite{jacques2023data}. We have successfully applied the nonparametric approach to both models and present results for the double-gyre model \cite{jiang1995multiple, jacques2023data}, which features more complex dynamics and a more intricate RC (Fig. \ref{fig:6}).

The committor is computed between the ``up'' ($q=1$) and ``down'' ($q=0$) steady states of the system's stream function. \cite{jacques2023data}. The steady-state probability as a function of the committor reveals, in addition to the ``up'' and ``down'' states, two marginally stable intermediate states located at $q \approx 0.44$ and $q\approx 0.58$ (Fig. \ref{fig:6}a). This is supported by Fig. \ref{fig:6}b, which shows the probability as a function of the committor and the model variable $A_1$, where one can clearly see the intermediate states. A segment of the trajectory (Fig. \ref{fig:6}b) illustrates complex dynamics: the system starts at the down state and goes to one intermediate state, then to the other, proceeds to the up state and then returns directly to the down state, suggesting the existence of multiple transition pathways. The high sensitivity of the nonparametric approach enables the direct identification of such low-populated intermediates. Accurate modeling of such transition and intermediate states is particularly important for ML approaches, aimed at accelerating rare event simulations in weather and climate systems \cite{lam2023learning,bi2023accurate}, as these states are often underrepresented in the training sets.
The validation criterion (Fig. \ref{fig:6}c) and the predicted versus observed committor plot (Fig. \ref{fig:6}d) confirm that the obtained RC closely approximates the committor. 

\subsection*{Committor for disease dynamics}
An optimal RC can be used to construct a theoretically optimal model of disease progression dynamics \cite{banushkina2025data}. We illustrate this by analyzing acute kidney injury (AKI), a sudden decline in kidney function often triggered by acute illness. AKI affects approximately 13.3 million people worldwide and causes an estimated 1.7 million deaths annually \cite{mehta2016recognition}. We focus on AKI stage 3 (AKI3), the most severe form. To illustrate the utility of incorporating histories, we use a single input variable: the time-series of serum creatinine (sCr), extracted from a highly irregular and imbalanced longitudinal clinical dataset (see Methods). 

\begin{figure}[h!]
	\centering
	\includegraphics[width=\linewidth]{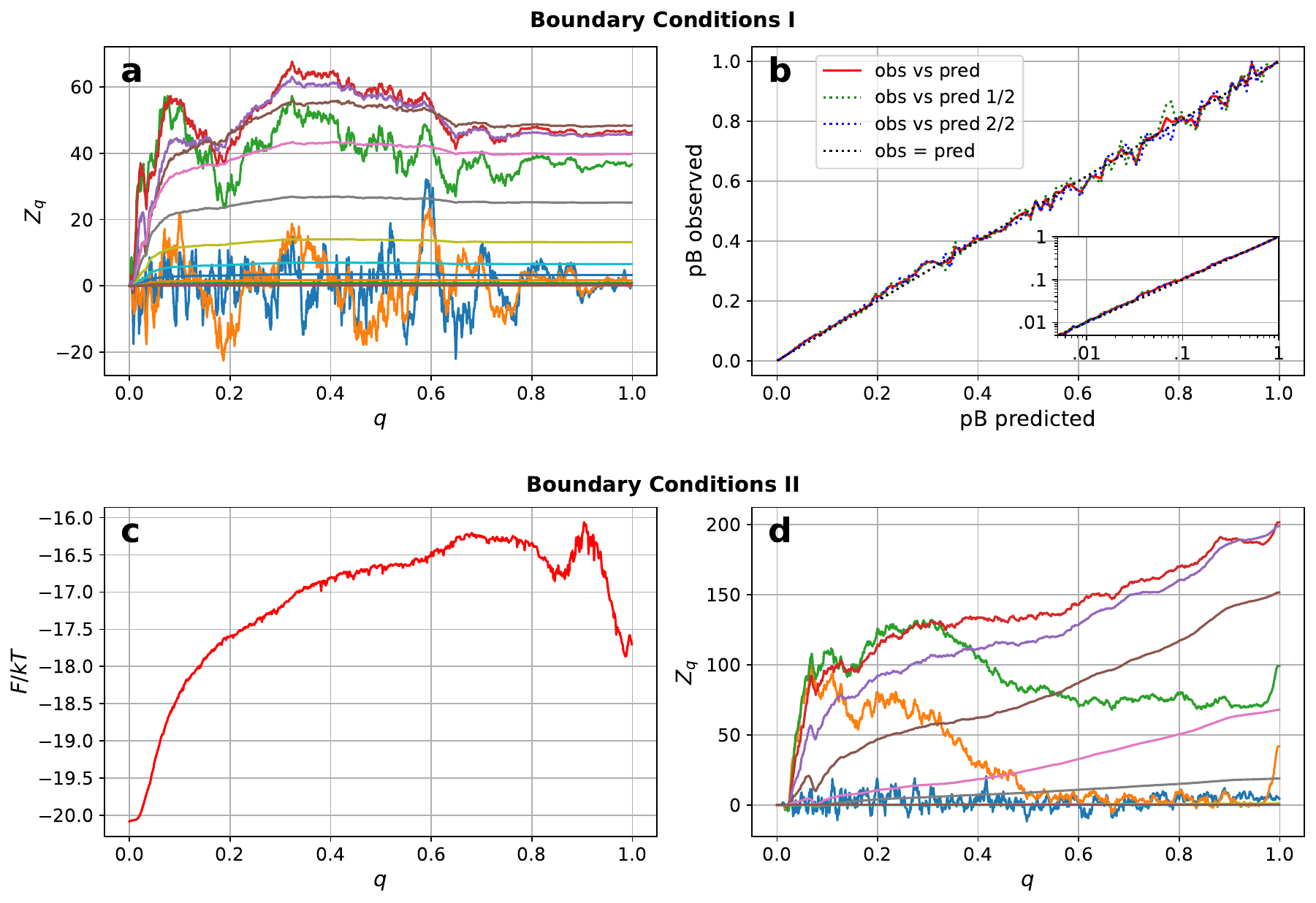}
	\caption{Committor optimization for AKI dynamics using a single sCr time-series. Boundary conditions I - likelihood of developing AKI3 during a hospital stay: (\textbf{a}) The validation criterion, $Z_q$, is relatively constant (except in the region of $q\sim 0$), deviations are comparable to statistical fluctuations; (\textbf{b}) Predicted and observed probabilities agree. The inset shows the comparison on a log-log scale. Boundary conditions II - dynamics between healthy and diseased (AKI3) states: (\textbf{c}) Equilibrated FEP describing the overall dynamics of AKI3 development. \textbf{d}) The validation criterion, $Z_q$, shows that RC is relatively close to optimal, though it could be further improved by incorporating additional input variables. In this scenario, in contrast to the first one, only a subset of trajectories reaches boundary states, resulting in biased observed probabilities (see Fig. \ref{fig:3}). Consequently, the predicted vs. observed plot is omitted.
	}
	\label{fig:7}
\end{figure}

We consider two scenarios for describing the dynamics, each defined by a different set of boundary conditions. In the first scenario (Figs. \ref{fig:7}a-b), each hospital admission is treated as an independent trajectory. Surrogate endpoints are defined as follows: if a patient experiences AKI3, identified using a standard clinical algorithm \cite{kdigo2012kdigo}, the trajectory ends in state $B$; otherwise, it ends in state $A$. This design ensures that every trajectory terminates in a boundary state, allowing accurate estimation of observed probabilities. The optimal RC in this scenario quantifies the likelihood of developing AKI3 during a hospital stay.
Figure \ref{fig:7}a shows that $Z_q$ remains relatively constant, except in the region of very small $q$. This indicates that the computed RC closely approximates the committor function, apart from that low-$q$ region. Predicted and observed probabilities (Fig. \ref{fig:7}b) are in strong agreement. In contrast, optimization without histories yields significantly worse results (Supplementary Fig. \ref{fig:S5}).

In the second scenario (Figs. \ref{fig:7}c-d), we aim to describe the dynamics between healthy and diseased states. While AKI3 can be specified using a standard clinical algorithm, no such algorithm exists for objectively defining the healthy state. To address this, we use results from the first scenario to construct a data-driven definition: points with a very low likelihood of developing AKI3 ($q<0.01$) are classified as healthy. Each patient's full clinical history is treated as a single trajectory. The validation criterion (Fig. \ref{fig:7}d) shows slightly larger deviations from constancy compared to Fig. \ref{fig:7}a particularly near $q \approx 0$. Both plots demonstrate that incorporating histories allows even a single input variable to approximate the optimal RC reasonably well. Additional input variables could further improve the RC. 

The determined RC predicts the likelihood of developing AKI3 significantly earlier than the standard clinical algorithm, which diagnosed AKI3 only after it has already occurred. This makes the RC well-suited for personalized monitoring  of patients conditions and early diagnostics. However, here we want to illustrate the utility of optimal RCs in the developing data driven models of disease dynamics, where a diffusive model along the optimal RC provides a clear, visually intuitive, and quantitatively accurate representation of overall disease dynamics \cite{banushkina2025data}. Fig. \ref{fig:7}c shows the equilibrated FEP, which describes the progression from the healthy to the AKI3 state. In particular, the rarity of the disease is explained by the activation barrier (at $q\sim 0.9$); once a patient crossed the barrier, the critical, rate-limiting step, they rapidly progress downhill to the disease state. Patients before, near the top and after the barrier can be examined to gain insight into the key stages of disease development. Interestingly, the FEP shows an intermediate state $q\sim 0.85$, indicating additional complexity in the disease dynamics.

\section*{Discussion}
In this work, we proposed a nonparametric approach for RC optimization that incorporates trajectory histories. The use of histories compensates for missing or incomplete information (such as a limited set of CVs), enhances robustness, and naturally avoids overfitting. The power of the approach was demonstrated on a challenging protein folding dataset across increasingly difficult tasks. The method consistently produced accurate committor estimates, as verified by the stringent validation criterion, and high-resolution FEPs which describe the system's dynamics at the temporal resolution of the trajectory. These results demonstrate that unbiased, brute-force atomistic simulations, when combined with rigorous analysis, can uncover a surprisingly intricate free energy landscape underlying the folding dynamics of a relatively small protein (see e.g., Fig. \ref{fig:4}f). Such a system provides an excellent and challenging benchmark for the development and evaluation of methods for RC and CV optimization, dynamical analysis, and enhanced sampling. The broad applicability of the framework was further illustrated by applying it to a diverse set of systems, including dynamics in phase space, a conceptual ocean circulation model, and a clinical longitudinal dataset.

A key strength of the proposed framework is its generality. Unlike many alternative methods, it does not rely on common assumptions such as detailed balance, equilibrium sampling, or overdamped dynamics. It is applicable to systems with missing variables, irregular time series, or incomplete observables, and to cases where exhaustive sampling is infeasible. In particular, it can be used to analyze systems governed by Fokker-Planck equations with position-dependent diffusion coefficients, underdamped Langevin dynamics, or rare event dynamics in exponentially large configuration spaces.

This generality also distinguishes our method from parametric approaches, which require the explicit definition of a functional form approximating the RC, typically a neural network architecture that is system specific and demands substantial domain knowledge. The nonparametric approach does not assume any functional form and can closely approximate arbitrary RCs. It requires minimal system-specific expertise, one only needs to provide the collective variables. Supplementary Fig. \ref{fig:S3} shows the basic Python script used in our analysis.

The framework emphasizes a shift in methodology: rather than learning a global RC function that generalizes across the entire configuration space, it focuses on computing optimal RCs only for the configurations that are actually sampled. This avoids the critical assumption that the training data must extensively sample the full configuration space, an assumption that often fails in rare-event scenarios or systems with exponentially large configuration spaces, where new configurations frequently fall out-of-distribution. This limitation also undermines the main advantage of parametric methods - their ability to rapidly evaluate RC values for unseen configurations. Importantly, many practical tasks, such as enhanced sampling, can be performed without knowing a global RC function. For example, one can progressively improve sampling by initiating new trajectories near the transition state, identified from the RC time-series \cite{krivov_nonparametric_2021}. When RC values are needed for new configurations, the RC can simply be re-optimized using newly available data, which also helps mitigate issues such as model drift, particularly in clinical applications.

The shift in methodology is supported by the proposed validation criterion $Z_q$, which assesses RC quality based on statistical independence across time scales. Specifically, RC optimization is conducted on the shortest timescale, while validation is performed across the full range of timescales. In contrast, standard approaches, which evaluate RCs based on train/test splits assume identically distributed data, a condition rarely satisfied in rare-event sampling. 

In comparison with GLE approaches, both the proposed approach and GLE approaches incorporate memory effects, but in fundamentally different ways. The proposed approach leverages memory, or more precisely the system's history, to determine a complex, nonlinear mapping from the history space to a position along the optimal RC. Once the position is determined, the dynamics are modeled using a simple diffusive model, yielding a clear and intuitive description of the system's overall dynamics. In contrast, GLE-based approaches employ a simple mapping to the RC while embedding all the dynamics complexity within the memory kernel. This results in non-Markovian dynamics that depend on the entire history of systems evolution. While the memory kernel formally encapsulates all relevant dynamical information, it often acts as a ``black box'', making mechanistic interpretation, such as identifying transition states, more challenging without explicit simulation. 

In summary, the proposed approach, combined with the stringent validation criterion,
provides a general, flexible, and robust framework for RC optimization in complex dynamical systems and longitudinal datasets, without relying on extensive sampling. 

\section*{Methods}

\subsection*{Nonparametric optimization with histories}
The proposed nonparametric framework is derived for a general discrete-time, finite-state Markov chain and does not rely on assumptions commonly made in alternative approaches \cite{krivov_nonparametric_2021}. While the theoretical derivation is based on a Markov chain formulation, the construction of such a chain is not required in practice. The method operates directly on the time series of a set of CVs, making it broadly applicable even when the underlying dynamics are only partially observed or irregularly sampled. Here we briefly summarize the approach, for details see Supplementary Information.

The nonparametric approach in the general setting of a non-equilibrium ensemble of trajectories with variable time intervals is as follows. All trajectories are assumed to be concatenated into a single long trajectory for analysis. Let $t_k$ denote time corresponding to frame $k$; under uniform sampling, this simplifies to $t_k=k\Delta t_0$, where $\Delta t_0$ is trajectory sampling interval. Starting with a seed RC time-series $r_0(t_k)$, satisfying the boundary conditions: $r_0(t_k)=0/1$ for $\vec{X}(t_k) \in A/B$ and $r_0(t_k)=0.5$ otherwise, the RC is iteratively improved. At the $m$-th step, the update is given by:
\begin{align*}
	r_{m+1}(t_k) &= r_m(t_k) + \delta r(t_k) \notag \\
	&= r_m(t_k) + \tilde{I}_b(t_k)\,g(t_k)\,f(y_m(t_k), r_m(t_k)),
\end{align*}
where $\tilde{I}_b(t_k)$ is a ``not at boundary'' indicator function (zero at boundaries and one otherwise), $g(t_k)$ is a common envelope function, $f(y,r)=\sum_{ij} a_{ij} y^ir^j$ is a small degree polynomial (degree 6 here). The variable $y_m$ denotes a randomly selected CV used to improve the RC. The envelope function is useful to improve optimization around free energy minima, which are exponentially compressed along an optimal RC \cite{krivov_protein_2018}. Here we used $g(t_k)=\sigma(\pm (r_m(t_k)-r_m(t_0))/d)$, where $\sigma(x)=1/(1+\exp(-x))$, $t_0$ is randomly selected time moment uniformly distributed over trajectory frames, and $d=0.001$.

The coefficients $a_{ij}$ that provide the best update for the committor RC are found as a solution of a linear equation, minimizing the following functional (see Supplementary Information)
$$\min_{a_{ij}}\sum_k (r_m(t_{k+1})-r_{m+1}(t_k))^2 I_t(t_k)+\gamma\delta r^2(t_k),$$ where indicator function $I_t(t_k)$ equals $1$ if frames $t_k$ and $t_{k+1}$ belong to the same trajectory and $0$ otherwise, ensuring that only transitions within the same trajectory are considered, and the regularization term with $\gamma$ is used to penalize large changes. Thus, each new iteration uses information contained in a new CV to improve the putative RC. If such an iterative process converges to the optimal RC, the set of CVs is called complete.
For an incomplete set of CVs, i.e., one which does not contain sufficient information to determine the optimal RC, one can use optimization with histories to compensate for missing information.

The optimization with histories is implemented by using the RC and CVs values at earlier times via the following variation 
$$r_{m+1}(t_k)=r_m(t_k)+\tilde{I}_b(t_k)g(t_k)f(y_m(t_l),r_m(t_l)),$$
where $l=k-\Delta_h$ and $\Delta_h>0$ is a randomly selected time delay drawn at each iteration from a given set. If the frames at times $t_k$ and $t_l$ belong to different trajectories, the index $l$ is replaced by $k_0$, the earliest frame from the same trajectory as frame $t_k$. This ensures that all variations were computed using frames from the same trajectory. 

While various forms of the variation function $f$ can be used, $f(y_m(t_l), r_m(t_l))$ is employed as the default due to its strong performance across diverse systems.  Alternative choices include $f(y_m(t_{k-1}), y_m(t_k))$ for the Langevin dynamics, and $f(\ln(t_k - t_l), r_m(t_l))$ or $f(\ln(t_k - t_l), y_m(t_l))$ for the clinical dataset, where accounting for variable time intervals is essential due to high temporal heterogeneity.

The optimization stops, for example, after a fixed number of iterations is reached, or when RC changes become sufficiently small.

\subsection*{Committor validation criterion $Z_q$}
Once a putative optimal RC has been computed, it must be validated to ensure it closely approximates the desired optimal RC. We have developed a stringent validation criterion for the committor, $Z_q$. This criterion is computed from the RC time-series $r(t_k)$ by numerically integrating the following derivative
$$\frac{d Z_q(x,\Delta)}{dx}=1/\Delta \sum_k \delta [x-r(t_k)][r(t_{k+\Delta})-r(t_k)]I_t(t_k),$$
where $\delta[\cdot]$ is the Dirac delta function and $\Delta$ is the observation time-scale or lag time in terms of trajectory frame indices, the number of frames separating two points in the time-series. For uniformly sampled trajectories, this corresponds to a physical time interval of $\Delta t = \Delta \cdot \Delta t_0$. 

Under uniform sampling, this expression computes the average displacement of the trajectory $\langle r(t+\Delta t)-r(t)\rangle$ conditioned on the trajectory starting from the point $r(t)=x$. For the committor function, this average displacement is zero for any lag time $\Delta$ and any $x$, except in regions involving transitions to or from the boundary states \cite{krivov_nonparametric_2021}. Thus, for an RC that closely approximates the committor $Z_q(x,\Delta)$ is constant. By employing the transition path segment summation scheme $Z_q(x,\Delta)$ can be made constant for all $x$ \cite{krivov_reaction_2013}. Specifically, all trajectories that visit boundary nodes are divided into segments that either end and/or start with boundary nodes, ensuring that no boundary node lies within a segment. To compute  $Z_q(x,\Delta)$ for a specific value of $\Delta$, each segment ending in a boundary state is extended by $\Delta$ steps using the last value. For example, for $\Delta=2$,  a trajectory segment ending in state $B$ as $[..., r(t_{k-2}), r(t_{k-1}), 1]$ is extended for 2 steps as  $[..., r(t_{k-2}), r(t_{k-1}), 1, 1, 1]$. This  procedure assumes that trajectories are of practically infinite length. For finite-length trajectories, the extended segment must not exceed the original trajectory length. In the case of variable sampling intervals, the criterion averages over the different intervals and remains constant for an optimal RC.

For equilibrium dynamics $Z_q$ equals to the cut-profile $Z_{C,1}$ \cite{krivov_nonparametric_2021, krivov_reaction_2013}. While for nonequilibrium sampling
the constant values of $Z_q$ for optimal RC may differ across different lag times $\Delta $, in equilibrium sampling all values of $Z_q$ and $Z_{C,1}$ are identical and independent of $\Delta $. For a sub-optimal RC (underfitting) $Z_{C,1}(r, \Delta )$ typically decreases with increasing $\Delta$, approaching $N_{AB}$ as $\Delta  \rightarrow \infty$. In contrast,  overfitting breaks this monotonic behavior. For example, $Z_{C,1}(r, 1)$ may become smaller than  $Z_{C,1}(r, k)$ for $k>1$, indicating that the RC is overfitted around $r$ \cite{krivov_reaction_2013, krivov_is_2010}.

For an RC, optimized at a time interval of $\Delta t$, cut-profile $Z_q(r, 1)$ should be identically constant. Deviations from this constant value arise due to statistical fluctuations and can serve as a measure of their magnitude.

\subsection*{Diffusion coefficient}
The diffusion coefficient along an RC can be accurately estimated, for relatively small $\Delta t$ and assuming uniform sampling, as \cite{krivov_reaction_2013}
$$D(r;\Delta t)=\frac{Z_{C,1}(r;\Delta t)}{\Delta t Z_H(r;\Delta t)}$$
where $Z_H(r;\Delta t)$ is the standard histogram density, computed using a time step $\Delta t$. The constancy of $Z_{C,1}$ for the committor means that the value of the diffusion coefficient is invariant across all time-scales $\Delta t$ as expected for Markovian dynamics.

Transformation to the natural coordinate $\tilde{q}$ with constant diffusion coefficient $D(\tilde{q})=1$ is performed by numerically integrating $d\tilde{q}=D(q)^{-1/2} dq$.

\subsection*{Predicted vs. observed plot} 
The plots are obtained by binning the putative RC coordinate into 1000 bins and computing statistics for all trajectory segments starting from the current bin and progressing towards the boundary states. For the committor RC, the fraction of trajectories ending in node $B$ is computed. For the MFPT RC, the mean time to node $B$ is computed. The plots are computed for the full trajectory as well as for the first and second halves to estimate statistical fluctuations.

It should be noted that, agreement in these plots does not confirm that the RC approximates the committor. Rather it indicates that the RC is properly ``calibrated'' \cite{Hond2022}. For example, a sub-optimal RC $r(t)$ can be transformed into a still sub-optimal RC $q(r(t))$, by training a ML classifier to predict the next boundary state ($0/1$), which will yield perfect agreement. In our analyses, the RC is optimized at the shortest time scale and is not trained to predict the next boundary. Therefore, agreement in the predicted vs. observed plot, which reflects dynamics at the longest time scale, indicates that the RC remains optimal across all time scales.

\subsection*{Reweighting factors}
For non-equilibrium sampling, re-weighting factors describe the necessary adjustments to the weight of each point to reproduce equilibrium sampling (for an equilibrium trajectory, these factors are equal to 1) \cite{krivov_nonparametric_2021}. Thus one can obtain an equilibrium FEP along the committor, which can be used
to determine important properties of the dynamics exactly.  For
a detailed description of re-weighting factors and the algorithm to compute them, see \cite{krivov_nonparametric_2021, krivov_optimal_2011}.

\subsection*{Realistic protein folding trajectory}
We analyzed a long equilibrium atomistic protein folding trajectory of the HP35 Nle/Nle double mutant (PDB ID: 2f4k \cite{kubelka2006sub}) consisting of 1509392 frames simulated at 380 K and saved at intervals of 0.2 ns \cite{piana_protein_2012}. The protein contains 577 atoms. The native basin (node $B$) is defined by a root-mean-square deviation (RMSD) from the native 2f4k structure of less than  1.0~\text{\AA}, while the denatured basin (node $A$) is defined by an RMSD greater than 10.5~\text{\AA}. The trajectory contains $2N_{AB}=146$ folding/unfolding transitions between these boundary states. In previous applications of the nonparametric approach \cite{krivov_protein_2018, krivov_blind_2020}, distances between randomly selected pairs of atoms ($i$ and $j$) were used as CVs, defined as $y(t)=\|\vec{r}_i(t)-\vec{r}_j(t)\|$. To demonstrate the generality and robustness of the approach, we instead use the sine and cosine of the backbone  $\phi$ and $\psi$ dihedral angles as CVs, with a total of 136 variables. 

The ensemble modeling clinical datasets was obtained by resampling the original trajectory as follows. Starting from a random point in the original trajectory, a short trajectory was extended with a 0.1 probability of stopping at each step. Additionally, each step had a 0.5 probability of being saved; and each dataset value had a 0.3 probability of being missed. The resulting ensemble contains $\sim 300000$ trajectories, averaging 5 steps in length, with 30\% of values missing.

\subsection*{Conceptual ocean models}
The dynamics of the reduced double-gyre model are governed by a nonlinear system of stochastic differential equations defined on four Fourier coefficients $A_i$ for $i=1,2,3,4$ \cite{jiang1995multiple}. The ocean models were simulated using the implementation and parameter settings provided in Ref.~\citenum{jacques2023data}.  A trajectory of $10^6$ frames was generated, containing 4680 transitions between the ``up'' and ``down'' states, as defined in Ref.~\citenum{jacques2023data}.

\subsection*{Longitudinal clinical dataset}
We analyzed a fully anonymized, retrospective longitudinal clinical dataset comprising all patients admitted to Leeds Teaching Hospitals NHS Trust between 2012 and 2021. The dataset includes 2,731,377 serum creatinine (sCr) measurements from 274,451 unique patients across 728,471 distinct hospital admissions. Among these, 70,879 patients (26\%) had only a single sCr record, while only 10,390 patients (3.8\%) experienced at least one episode of AKI3. The time intervals between consecutive sCr tests varied widely, ranging from hours to several years, with a median interval of approximately 24 hours.

\subsection*{Data Availability}
The datasets generated and analyzed during this study are available from the corresponding author upon reasonable request. Hospital data used in this work are subject to patient privacy regulations and cannot be shared publicly; access to these data should be requested from Leeds Teaching Hospitals NHS Trust. The protein folding trajectory should be requested from \href{https://www.deshawresearch.com/resources.html}{D. E. Shaw Research}.

\subsection*{Code Availability}
The Python library implementing the framework described in this study is available at \href{https://github.com/krivovsv/optimalrcs}{https://github.com/krivovsv/optimalrcs}. Analysis notebooks  are available at
\href{https://github.com/krivovsv/histories}{https://github.com/krivovsv/histories}.

\subsection*{Author Contribution}
S.V.K. conceptualized the study and developed the methodology. S.V.K. and P.V.B. analyzed the datasets, interpreted the results, and wrote the manuscript.

\subsection*{Competing Interests}
The authors declare no competing interests.

\section*{Acknowledgments}
The work is partially supported by Kidney Research UK (KRUK) award number (RP\_018\_20230628).  The authors are grateful to  D. E. Shaw Research for providing access to the protein folding trajectory.


%

\end{document}


\title{Supplementary Information for paper ``Nonparametric Reaction Coordinate Optimization with Histories for Rare Event Dynamics''}
\author{Polina V. Banushkina, Sergei V. Krivov}
\affiliation{Astbury Center for Structural Molecular Biology, Faculty of Biological Sciences, University of Leeds, Leeds LS2 9JT, United Kingdom}

\maketitle

\setcounter{figure}{0}
\renewcommand{\thefigure}{\arabic{figure}}
\renewcommand{\figurename}{Supplementary Figure}

\section{Theory}
For the reader's convenience, this section summarizes the derivation of equations within the nonparametric framework for reaction coordinate (RC) optimization, covering both the committor function and the mean first passage time (MFPT) \cite{krivov_reaction_2013, banushkina_nonparametric_2015, banushkina_optimal_2016, krivov_protein_2018, krivov_nonparametric_2021, banushkina2025data}. The derivation proceeds in three steps. First, we derive the nonparametric optimization approach for a finite-state Markov chain. Next, we reformulate the optimization functional in terms of RC time-series, which removes the need to explicitly construct the Markov chain. Finally, we obtain the equations for optimal variation parameters using a basis function expansion.

While the theoretical foundation relies on a Markov chain model, the construction of
such a chain is not required in practice. The approach operates directly on time-series of collective variables (CVs). This makes the method broadly applicable, even when the underlying dynamics are only partially observed or irregularly sampled.

We consider stochastic dynamics described by an ergodic, discrete-time, finite-state Markov chain with transition probability matrix $P_\tau(i|j)$, representing the probability of transitioning from state $j$ to state $i$ after a time interval $\tau$. This formulation is general and does not rely on assumptions commonly made in alternative approaches. It does not require detailed balance, equilibrium sampling, or constant diffusion tensors. In particular, it can be rigorously applied to systems governed by arbitrary Fokker-Planck equations with state-dependent diffusion tensors, and to underdamped Langevin dynamics.

The committor function $q(j)$, which gives the probability that a trajectory initiated at state $j$ reaches state $B$ before state $A$, satisfies the following equation for all intermediate states $j \notin \{A, B\}$:
\begin{equation}
	\label{eq:q}
	\sum_i P_\tau(i|j) [q(i) - q(j)] = 0,
\end{equation}
with boundary conditions $q(A) = 0$ and $q(B) = 1$.

\subsection{Equilibrium sampling}
Consider a long equilibrium trajectory that samples the configuration space of the finite-state Markov chain. Let $n(i|j)$ denote the number of observed transitions from state $j$ to state $i$, and let $N$ be the total number of transitions in the trajectory. Under equilibrium conditions, the transition counts satisfy detailed balance: $n(i|j) = n(j|i)$. Assuming ergodicity and stationarity, we approximate the transition counts as $n(i|j) \approx P_\tau(i|j) P^{\text{eq}}(j) N$, where $P^{\text{eq}}(j)$ is the equilibrium probability of state $j$.

Under these conditions, the committor function $q(j)$ can be obtained by minimizing the following quadratic functional
\begin{equation}
	\label{eq:func_nij}
	\min_{r(j)} \sum_{ij} n(i|j)\left[r(i)-r(j)\right]^2,
\end{equation}
subject to boundary conditions $r(A)=0$ and $r(B)=1$. Here, $r(j)$ is a putative reaction coordinate intended to approximate the committor. Taking the derivative with respect to $r(j)$ and applying the detailed balance condition, one recovers Eq.~\ref{eq:q}. 

For trajectories sampled at uniform time intervals $\Delta t_0$, the optimization functional (Eq.~\ref{eq:func_nij}), denoted as $\Delta r^2$, can be expressed directly in terms of the RC time-series
\begin{equation}
	\label{eq:func_dt0}
	\Delta r^2 = \sum_{k=1}^{N-1} \left[r(k\Delta t_0 + \Delta t_0) - r(k\Delta t_0)\right]^2,
\end{equation}
where $r(k\Delta t_0)$ denotes the value of the RC time-series at time $k\Delta t_0$. When the RC equals the committor function, the optimization functional $\Delta r^2$ attains its minimum value of $2N_{AB}$ \cite{krivov_reaction_2013}. Here $N_{AB}$ is the number of transitions from state $A$ to state $B$ in the equilibrium trajectory, $N_{AB} = N_{BA}$.

To find the RC that minimizes this functional, we employ an iterative optimization procedure. The seed RC time-series, denoted $r_0(k\Delta t_0)$, is defined to satisfy the boundary conditions as follows: $r_0(k\Delta t_0) = 0$ for all frames belonging to state $A$, $r_0(k\Delta t_0) = 1$ for all frames in state $B$, and $r_0(k\Delta t_0) = 1/2$ for all other frames. At each iteration $m$, the RC is updated according to
\begin{equation}
	\label{eq:var3}
	r_{m+1}(k\Delta t_0) = r_m(k\Delta t_0) + \delta r(k\Delta t_0)
\end{equation}
where the variation $\delta r(k\Delta t_0)$ assumed to be zero at the boundaries, in order to preserve the constraints.

Let the variation be expressed as a linear combination of $L$ basis functions $f_l(t)$
\begin{equation}
	\label{eq:var}
	\delta r(t) = \sum_{l=1}^L \alpha_l f_l(t),
\end{equation}
where each basis function satisfies the boundary conditions $f_l(A) = f_l(B) = 0$.

The optimal coefficients, which yield the best approximation to the committor function for the considered variation, are obtained by minimizing the $\Delta r^2$ functional (Eq.~\ref{eq:func_dt0}). These coefficients are found by setting the derivative of the functional with respect to each $\alpha_l$ to zero, resulting in the following system of linear equations
\begin{equation}
	\sum_{k=1}^{N-1} [r_{m+1}(k\Delta t_0 + \Delta t_0) - r_{m+1}(k\Delta t_0)] f_l(k\Delta t_0) = 0,
\end{equation}
for $l = 1, \dots, L$; where $f_l(k\Delta t_0)$ denotes the $l$-th basis function evaluated at time $k\Delta t_0$, and $r_{m+1}$ is expressed using the basis expansion defined in Eqs. \ref{eq:var3} and \ref{eq:var}.

Since the functional decreases with each iteration and is bounded from below, the optimization procedure converges. In the limit of extensive sampling and in absence of overfitting, convergence to the theoretical lower bound of $2N_{AB}$, indicates that the RC closely approximates the committor \cite{krivov_reaction_2013}.

To assess the quality of the resulting RC one may apply the $Z_{C,1}$ validation criterion, described below.

\subsection{Non-equilibrium Sampling Using Ensembles of Trajectories}
Consider now the case of non-equilibrium sampling, where instead of a single long equilibrium trajectory, we have an ensemble of trajectories initiated from arbitrary states of a Markov chain. Let $n(i|j)$ denote the aggregated number of observed transitions from state $j$ to state $i$, and let $n_j$ be the number of times state $j$ was visited across the ensemble. Under this setting, the transition counts can be approximated as $n(i|j) \approx P_\tau(i|j) n_j$. Let $r(k)$ denote the reaction coordinate evaluated at state $k$, and $\delta r(k)$ its variation.

Since the trajectories are not sampled from equilibrium, the detailed balance condition no longer holds: $n(i|j) \ne n(j|i)$, and the optimization functional must be modified
\begin{equation}
	\label{eq:func_nij_ne}
	\min_{\delta r} \sum_{i,j} n(i|j) [r_m(i) - r_{m+1}(j)]^2
\end{equation}

Taking the derivative of the functional with respect to the variation $\delta r(k)$ yields the condition for the optimal update
\begin{equation}
	\sum_i P_\tau(i|k) [r_m(i) - r_{m+1}(k)] = 0
\end{equation}

When the process converges, i.e., $r_{m+1}(k) = r_m(k) = q(k)$, Eq.~\ref{eq:q} is recovered.

To express the optimization functional in terms of the RC time-series, we assume that all trajectories in the ensemble are concatenated into a single long trajectory. Let $r(k\Delta t_0)$ denote the RC sampled at uniform time intervals $\Delta t_0$. Under this representation, the functional in Eq.~\ref{eq:func_nij_ne} becomes
\begin{equation}
	\label{eq:func_dt0_ne}
	\min_{\delta r} \sum_{k=1}^{N-1} \left[r_m(k\Delta t_0 + \Delta t_0) - r_{m+1}(k\Delta t_0)\right]^2 I_t(k\Delta t_0) 
\end{equation}
where, $I_t(k\Delta t_0) = 1$ if both $k\Delta t_0$ and $k\Delta t_0+\Delta t_0$ belong to the same trajectory, and $0$ otherwise. 

\subsection{Generalization to Variable Sampling Intervals}
\label{sect:var}
To justify the use of this functional for trajectories with variable sampling intervals, consider the committor equation (Eq.~\ref{eq:q}) evaluated over a doubled time interval
\begin{multline}
	\sum_i P_{2\tau}(i|j) [q(i) - q(j)] = \\
	\sum_i \sum_k P_{\tau}(i|k) P_{\tau}(k|j) [q(i) - q(k)] + \\
	\sum_i \sum_k P_{\tau}(i|k) P_{\tau}(k|j) [q(k) - q(j)] = 0
\end{multline}
This identity holds for all intermediate states $j$, sufficiently far from the boundaries (i.e., that both $j,k \notin \{A, B\}$). Thus, neglecting errors near the boundaries, the committor satisfies the same equation regardless of the sampling interval. This implies that the optimization functional can be extended to trajectories with variable sampling intervals.

The functional, generalized to trajectories with variable sampling intervals and including a regularization term, can be written as
\begin{equation}
	\label{eq:func_final}
	\min_{\delta r} \sum_{k=1}^{N-1} [r_m(t_{k+1}) - r_{m+1}(t_k)]^2 I_t(t_k) + \gamma \sum_{k=1}^N \delta r^2(t_k),
\end{equation}
where $t_k$ denotes the time associated with frame $k$, and $r(t_k)$ is the value of the RC time-series at that time. The indicator function $I_t(t_k) = 1$ if both $t_k$ and $t_{k+1}$ belong to the same trajectory, and $0$ otherwise. The regularization term, weighted by $\gamma$, penalizes large variations and improves robustness, particularly when incorporating trajectory histories, as discussed below.

Assuming the variation is expressed as a linear combination of basis functions (Eq.~\ref{eq:var}), the optimal coefficients $\alpha_l$ are found by setting the derivative of the functional to zero
\begin{multline} 
\sum_{k=1}^{N-1} \left[r_m(t_{k+1}) - r_{m+1}(t_k)\right] I_t(t_k) f_l(t_k) + \\ \gamma \sum_{k=1}^N \delta r(t_k) f_l(t_k) = 0, \ \text{for } l = 1, \dots, L.
\end{multline}

To accelerate convergence, $r_m$ may be substituted with $r_{m+1}$ in the first term, yielding a linear system for $\alpha_l$
\begin{multline}
	\sum_{k=1}^{N-1} [r_m(t_{k+1}) - r_m(t_k) + \delta r(t_{k+1}) - \delta r(t_k)] I_t(t_k) f_l(t_k) + \\
	\gamma \sum_{k=1}^N \delta r(t_k) f_l(t_k) = 0, \ \text{for } l = 1, \dots, L,
\end{multline}
 and $\delta r(t_k)$ is expressed using the basis expansion defined in Eq.~\ref{eq:var}.

\subsection{Mean First Passage Time}
The mean first passage time (MFPT) to state $A$ satisfies the following equation for a discrete-time finite Markov chain
\begin{equation}
	\label{eq:mfpt}
	\sum_i P_\tau(i|j) \left[\text{mfpt}(i) - \text{mfpt}(j) + \tau\right] = 0 \quad \text{for } j \notin A,
\end{equation}
with boundary condition $\text{mfpt}(A) = 0$. Here $\text{mfpt}(j)$ denotes the MFPT as a function of state $j$.

The corresponding optimization functional for  an ensemble of trajectories with variable sampling intervals is
\begin{multline}
	\label{eq:mfpt_func}
	\min_{\delta r} \sum_{k=1}^{N-1} \left[r_m(t_{k+1}) - r_{m+1}(t_k) + t_{k+1} - t_k\right]^2 I_t(t_k) \\
	+ \gamma \sum_{k=1}^N \delta r^2(t_k),
\end{multline}
where $t_k$ denotes the time associated with frame $k$, and $r(t_k)$ and $\delta r(t_k)$ represent the RC and its variation evaluated at time $t_k$, respectively. The additional term $t_{k+1} - t_k$ accounts for the elapsed time between frames, reflecting the MFPT structure.

Assuming the variation is expressed as a linear combination of basis functions (Eq.~\ref{eq:var}), the optimal coefficients $\alpha_l$ are found by solving the following linear system
\begin{multline}
	\sum_{k=1}^{N-1} [r_m(t_{k+1}) - r_m(t_k) + \delta r(t_{k+1}) - \delta r(t_k) + t_{k+1} - t_k] \\
	\times I_t(t_k) f_l(t_k) + \gamma \sum_{k=1}^N\delta r(t_k) f_l(t_k) = 0,
\end{multline}
  for $l = 1, \dots, L$, and $\delta r(t_k)$ is expressed using the basis expansion defined in Eq.~\ref{eq:var}.

\subsection{Variations}
The RC variation for the optimization without histories, and with variable sampling intervals, is taken as
\begin{equation}
	\label{eq:var2}
	r_{m+1}(t_k)=r_m(t_k) + \tilde{I}_b(t_k)g(t_k)f(y_m(t_k),r_m(t_k)),
\end{equation}
where $\tilde{I}_b(t_k)$ is a "not-at-boundary" indicator function, equal to $0$ at boundary frames and $1$ otherwise, ensuring that the variation vanishes at boundaries; $y_m(t_k)$ is a CV randomly selected at each iteration to improve the RC; $g(t_k)$ is a common envelope function used to enhance uniformity of optimization near free energy minima, which are exponentially compressed along an optimal RC.  We used $g(t_k)=\sigma(\pm (r_m(t_k)-r_m(t_0))/d)$, where $\sigma(x)=1/(1+\exp(-x))$, $t_0$ is randomly selected frame uniformly distributed over trajectory frames, and $d=0.001$.

$f(y, r)$ is a polynomial basis (of total degree 6 here)
\begin{equation}
	\label{eq:fyr}
	f(y,r)=\sum_{0\le i,j}^{i+j\le 6} a_{ij} y^i r^j
\end{equation}
where $a_{ij}$ are the polynomial coefficients, playing the role of $\alpha_l$ in Eq.~\ref{eq:var}, and each monomial $y^i r^j$ corresponds to a basis function $f_l$.

To incorporate histories, the variation is modified to use RC and CV values from earlier times 
\begin{equation}
	r_{m+1}(t_k)=r_m(t_k)+\tilde{I}_b(t_k)g(t_k)f(y_m(t_l),r_m(t_l)),
	\label{eq:varh}
\end{equation}
where $l=k-\Delta_h$ with $\Delta_h>0$. If the frames at times $t_{k}$ and $t_l$ belong to different trajectories, the index $l=k-\Delta_h$ is replaced by $k_0$, the earliest frame from the same trajectory as frame $t_k$. This ensures that all variations are computed using frames from the same trajectory.
During optimization, $\Delta_h$ is randomly selected from a predefined  set at each iteration. If $\Delta_h=0$, the standard variation with $f(y_m(t_k),r_m(t_k))$ is used.

While various basis functions can be employed, the polynomial from $f(y_m(t_l),r_m(t_l))$ has shown the best performance across many systems and is considered the default. Additional basis functions used include: $f(y_m(t_{k-1}),y_m(t_k))$ for the Langevin dynamics; and $f(\ln (t_k-t_l), r_m(t_l))$, and $f(\ln (t_k-t_l), y_m(t_l))$ for the clinical dataset, where accounting for variable time intervals is critical due to high temporal heterogeneity.

\subsection{Optimal RC  Validation Criteria}
We begin by formulating the committor validation criterion for a finite-state Markov chain.
Consider a non-equilibrium ensemble of trajectories sampled at interval $\tau$, described by a discrete-time Markov chain. Let $n(i|j)$ denote the number of observed transitions from state $j$ to state $i$, and let $r(j)$ be the value of the reaction coordinate at state $j$. The committor validation criterion $Z_q(x, \tau)$ is defined such that its derivative is
\begin{equation}
	\label{eq:zq_nij}
\frac{d Z_q(x,\tau)}{dx}=\sum_{ij} \delta [x-r(j)]n(i|j)[r(i)-r(j)],
\end{equation}
where $\delta[\cdot]$ is the Dirac delta function. Summing over $i$ and invoking the committor equation (Eq.~\ref{eq:q}), one finds that the derivative vanishes, implying that $Z_q(x, \tau)$ is constant along the committor.  

The criterion, rewritten for RC time-series with uniform sampling interval $\Delta t_0$ is
\begin{multline}
	\label{eq:zq_t0}
	\frac{d Z_q(x,\Delta t_0)}{dx} = \sum_k \delta[x - r(k\Delta t_0)] \\ 
	\times  [r(k\Delta t_0 + \Delta t_0) - r(k\Delta t_0)]	I_t(k\Delta t_0)
\end{multline}
Since the committor satisfies the same equation for any multiple of the base sampling interval $\Delta t = n \Delta t_0$ (see Sect. \ref{sect:var}), the criterion generalizes to
\begin{multline}
	\label{eq:zq_t0}
	\frac{d Z_q(x,\Delta t)}{dx} = 1/n \sum_k \delta[x - r(k\Delta t_0)] \\ 
	\times  [r(k\Delta t_0 + \Delta t) - r(k\Delta t_0)]	I_t(k\Delta t_0)
\end{multline}
which remains constant for the committor RC, except near the boundaries.

To ensure constancy near boundaries, we employ the transition path segment summation scheme  \cite{krivov_reaction_2013}. Trajectories that visit boundary states are divided into segments that start and/or end at boundaries, ensuring that no boundary state lies within a segment. Each segment ending at a boundary is extended by $\Delta$ steps using the final value. 
For instance, for $\Delta=2\Delta t_0$,  a trajectory segment ending in state $B$ as $[..., r(k\Delta t_0-\Delta t_0), r(k \Delta t_0), 1]$ is extended for 2 steps as  $[..., r(k\Delta t_0-\Delta t_0), r(k \Delta t_0), 1, 1, 1]$. This  procedure assumes that trajectories are of practically infinite length. For finite-length trajectories, the extended segment must not exceed the original trajectory length. In sampling scenarios where trajectories terminate upon reaching a boundary, i.e., when the boundary conditions act as absorbing states (or traps), the validation criterion $Z_q$ remains constant only if trajectory lengths are made consistent across the ensemble. Specifically, segments ending at boundaries must be extended to match the lengths of trajectories that do not reach a boundary \cite{banushkina2025data}.
 
Trajectories with variable sampling interval can be considered as the superposition of trajectories with uniform sampling intervals, and thus have constant $Z_q$. For an ensemble of trajectories with variable sampling, the criterion takes the form
\begin{equation}
	\frac{d Z_q(x,\Delta)}{dx}=1/\Delta \sum_k \delta [x-r(t_k)][r(t_{k+\Delta})-r(t_k)]I_t(t_k),
\end{equation}
here $\Delta$ is the observation time-scale or lag time in terms of trajectory frame indices, the number of frames separating two points in the time-series. For uniformly sampled trajectories, this corresponds to a physical time interval of $\Delta t = \Delta \cdot \Delta t_0$. 

For an equilibrium trajectory, $Z_q$ coincides with $Z_{C,1}$ criterion, defined as 
\begin{equation}
	Z_{C,1}(x,\Delta t)=1/2 \sum^{x'}_k [r(k\Delta t+\Delta t)-r(k\Delta t)],
\end{equation}
where $\sum^{x'}_k$ denotes the sum over all $k$ that $x$ lies between $r(k\Delta t +\Delta t)$ and $r(k\Delta t)$ \cite{krivov_reaction_2013}. For nonequilibrium sampling the constant values of $Z_q$ for committor RC may differ across different lag times $\Delta t$. In contrast, in equilibrium sampling all values of $Z_q$ and $Z_{C,1}$ are identical and independent of $\Delta t$ and equal to $N_{AB}$ (up to statistical fluctuations). 

We now define a similar criterion for validating the mean first passage time (MFPT) RC.
The MFPT validation criterion $Z_\tau$, written for a non-equilibrium trajectory sampling Markov chain with sampling interval $\tau$, is
\begin{equation}
	\label{eq:zq_nij}
	\frac{d Z_q(x,\tau)}{dx}=\sum_{ij} \delta [x-r(j)]n(i|j)[r(i)-r(j)+\tau].
\end{equation}
Again, summing over $i$ and using the MFPT equation (Eq.~\ref{eq:mfpt}) shows that $Z_\tau(x, \tau)$ is constant when RC closely approximate the MFPT.

Following similar steps, one arrives at the validation criterion for the MFPT, $Z_\tau$, for non-equilibrium trajectories with variable sampling. It is computed from the RC time-series as
\begin{multline} 
\frac{d Z_\tau(x,\Delta)}{dx} = 1/\Delta \sum_k \delta[x - r(t_k)] \\ \times [r(t_{k+\Delta}) - r(t_k) + t_{k+\Delta} - t_k] I_t(t_k)
\end{multline}

\section{Analyses}
\subsection{RC optimization without histories for a complete set of CVs}
\label{sect:complete}
To demonstrate the generality and robustness of the approach, we consider a set of 136 CVs, consisting of the sine and cosine of backbone $\phi$ and $\psi$ dihedral angles. This set of CVs is not complete, meaning the nonparametric equilibrium optimization without histories cannot reach the theoretical lower bound of $\Delta r^2\sim 2N_{AB}$, as demonstrated in section \ref{sect:incomp}. It implies that variation using these CVs do not improve the putative RC further.

One strategy to make the set of CVs complete, is to use variations based on randomly selected pairs of CVs, e.g., $\delta r = f(y(t), z(t), r(t))$. A simpler alternative is to retain single-CV variations but expand the CV set to include combinations such as $y(t) \pm z(t)$. However, as shown here, the usage of such a complete CV set generally leads to overfitting.

Specifically, the following CVs are considered $y(t)=a_1x_{i_1}(t)+a_2x_{i_2}(t)+a_3x_{i_3}(t)$, where $a_i$ are chosen randomly from $\pm1$, while  $x_{i_1}$, $x_{i_2}$, $x_{i_3}$ are randomly chosen sine or cosine of backbone dihedral angles of nearby residues: $|res(i_1)-res(i_2)|\le 1$ and $|res(i_2)-res(i_3)|\le 1$, where function $res$ returns the residue number. 

Optimization was performed using the equilibrium nonparametric approach for uniformly sampled trajectories, with the functional defined in Eq.~\ref{eq:func_dt0} and variations defined by Eq.~\ref{eq:var2}. The procedure was terminated once $\Delta r^2 < 2N_{AB} = 146$, which required approximately 100{,}000 iterations.

\begin{figure}[h!]
	\centering
	\includegraphics[width=\linewidth]{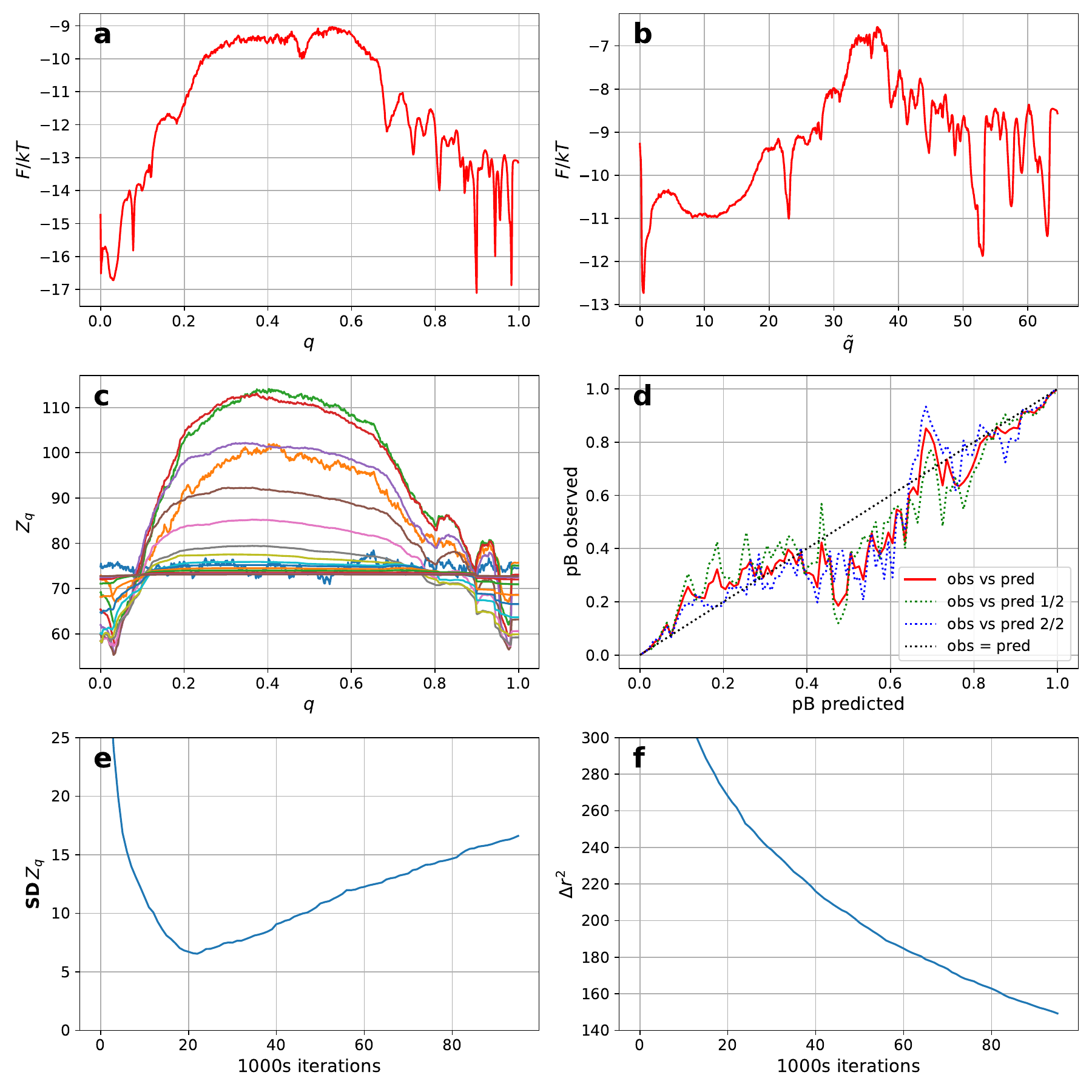}
	\caption{RC optimization without histories using a complete set of CVs. (\textbf{a}) Free energy as a function of the committor $q$. (\textbf{b}) Free energy as a function of the natural committor $\tilde{q}$, where $D(\tilde{q})=\text{const}$. (\textbf{c}) Validation criterion ($Z_q$) shows significant systematic variability, indicating that the putative RC does not closely approximate the committor and, in fact, overfits around the TS region. (\textbf{d}) Predicted and observed committors; (\textbf{e}) standard deviation of $Z_q$  and (\textbf{f}) $\Delta r^2$ functional as functions of iteration number.
}
	\label{fig:S1}
\end{figure}

Supplementary Fig. \ref{fig:S1} shows the results of optimization.  Supplementary Fig. \ref{fig:S1}c shows
that validation criterion is not constant and exhibit significant systematic variability, indicating that the obtained RC does not approximate the committor accurately. In particular, the TS region is overfitted, as $Z_q$, which for equilibrium trajectory equals to $Z_{C,1}$, does not decrease monotonically, i.e., $Z_{C,1}(r,1)<Z_{C,1}(r,\Delta)$ for $1 < \Delta$, which is an indication of overfitting \cite{krivov_protein_2018}. Additionally, the FEP around the TS (Supplementary Fig. \ref{fig:S1}a) is approximately 1 kT higher than that in Fig. \ref{fig:1}a. As a result of overfitting, the observed and predicted committor show poor agreement (Supplementary Fig. \ref{fig:S1}d).

To assess whether $Z_q$ deviates from a constant value, one can visually inspect the profiles, or use a descriptive cumulative characteristics, such as maximal standard deviation \textit{SD} of $Z_q(r, k)$. This involves computing the standard deviation for each $k$ and then determining the maximum. A small \textit{SD} indicates that $Z_q$ is approximately constant, indicating an optimal RC, whereas a large \textit{SD} suggests that the coordinate is not optimal.
Supplementary Fig. \ref{fig:S1}e, which measures the maximal standard deviation of $Z_q$ profiles shows that at the initial stage of the optimization \textit{SD} $Z_q$ was getting smaller, indicating improving approximation of the committor by the RC. However, after 20000 iterations, \textit{SD} $Z_q$ began to increase, while $\Delta r^2$  continued to decrease monotonically, indicating overfitting.

\subsection{RC optimization with histories for a complete set of CVs}
Optimization with histories was performed using the same set of CVs described in Section~\ref{sect:complete}, defined as:
$$
y(t) = a_1 x_{i_1}(t) + a_2 x_{i_2}(t) + a_3 x_{i_3}(t),
$$
where $a_i \in \{\pm 1\}$ are randomly selected coefficients, and $x_{i_1}$, $x_{i_2}$, $x_{i_3}$ are randomly chosen sine or cosine values of backbone dihedral angles from nearby residues.

Optimization was carried out using the nonparametric approach, with the functional defined in Eq.~\ref{eq:func_final} and variations defined by Eq.~\ref{eq:varh}. The history offsets were selected from the set $\Delta_h = \{0, 1, 2, 3, 4, 5\}$, and the regularization parameter was set to $\gamma = 0.2$. Ten optimization runs were performed to ensure robustness, each initialized with a different random seed and run for 100,000 iterations.

\subsection{RC optimization without histories for an incomplete set of CVs}
\label{sect:incomp}
In this section we show that equilibrium nonparametric optimization without histories, when applied to an incomplete set of CVs, even after extensive iterations, does not reach the theoretical lower bound of $\Delta r^2=2N_{AB}$, and eventually begins to overfit. The following CVs are considered $y(t)=x_{i}(t)$,  where  $x_{i}(t)$ is a randomly chosen sine or cosine of a backbone dihedral angle. The other parameters are the same as in Section \ref{sect:complete}.  Optimization was continued for 1,000,000 iterations, however the final values of $\Delta r^2 \approx 280 >2N_{AB}=146$, indicating that the putative RC is not fully optimized. 

\begin{figure}[h!]
	\centering
	\includegraphics[width=\linewidth]{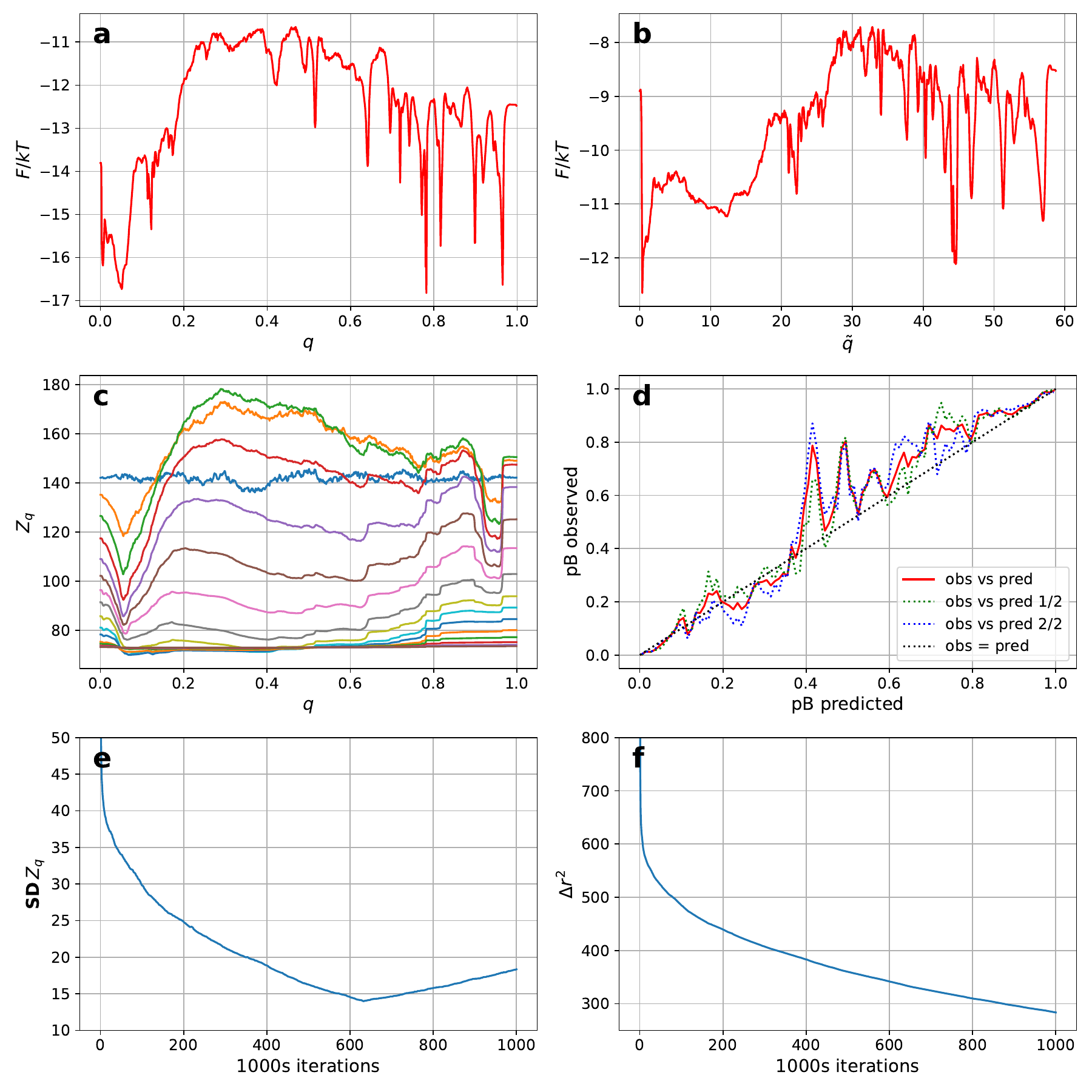}
	\caption{RC optimization without histories using an incomplete set of CVs. Notations as in Fig. \ref{fig:S2}.}
	\label{fig:S2}
\end{figure}  

Supplementary Fig. \ref{fig:S2} shows the results of the optimization. FEP (Supplementary Fig. \ref{fig:S2} a) is slightly lower (by $\sim 1$ kT) in the TS region than that at Fig. \ref{fig:1}a confirming that the RC is sub-optimal. The optimality criterion $Z_q$, which equals to $Z_{C,1}$ for the equilibrium trajectory, also indicates that the coordinate is sub-optimal, as $Z_{C,1}(q,1)$ is nearly twice as large as $Z_{C,1}(q,2^{16})$, whereas for committor all $Z_{C,1}$ are equal. The criterion further shows that the optimization begins to overfit the RC in the TS region as $Z_{C,1}(q,1)$ (blue) is lower than $Z_{C,1}(q,\Delta)$ for $\Delta=2$ (green), $\Delta=4$ (yellow). As the result, the predicted and observed committors show poor agreement. The maximal SD of $Z_q$ reaches a minimum at $\approx 600000$ iterations and then begins to increase (Supplementary Fig. \ref{fig:S2}e), confirming the onset of overfitting. Although $\Delta r^2$ continues to decrease monotonically, the rate of decrease slows, suggesting that even if the lower bound of $2N_{AB}$ can be reached, it may require many more millions of iterations.

\subsection{RC optimization with histories for an incomplete set of CVs}
The following CVs are considered $y(t)=x_{i}(t)$,  where  $x_{i}(t)$ is a randomly chosen sine or cosine of a backbone dihedral angle. Optimization used the functional defined in Eq.~\ref{eq:func_final} and variations defined by Eq.~\ref{eq:varh}, where $\Delta_h=\{0,1,2,3,4,5,6,7,8,9,10\}$, and with $\gamma=0.2$. Ten optimization runs were performed to ensure robustness, each initialized with a different random seed and run for 100,000 iterations.

\begin{figure}[h!]
	\centering
	\includegraphics[width=\linewidth]{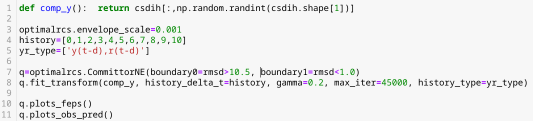}
	\caption{A basic Python script for RC optimization with histories using an incomplete set of CVs.}
	\label{fig:S3}
\end{figure}

Supplementary Fig. \ref{fig:S3} shows a basic Python script that performs the RC optimization, illustrating the simplicity of the nonparametric approach: one only needs to define a custom \texttt{comp\_y()} function that returns randomly chosen CVs.

\subsection{RC optimization with histories for highly irregular dataset}
The highly irregular ensemble of trajectories was obtained by resampling the original trajectory as follows: starting from a random point in the original trajectory, a short trajectory was extended with a 0.1 probability of stopping at each step. Each step had a 0.5 probability of being saved; and each value had a 0.3 probability of being missed. The resulting ensemble contains $\sim 300000$ trajectories, averaging 5 steps in length, with 30\% of s missing. Missing values were then imputed using the last known values in the same trajectory (all values are present at the first step in a trajectory). Optimization was performed using a complete set of CVs, as described in Section \ref{sect:complete}, using the functional defined in Eq.~\ref{eq:func_final} and variations defined by Eq.~\ref{eq:varh}, with $\Delta_h=\{0,1,2\}$ and $\gamma=0.02$.
Ten optimization runs were performed to ensure robustness, each initialized with a different random seed and run for 100,000 iterations.

\subsection{RC optimization with histories using a single CV}
The time-series of the RMSD from the native 2f4k pdb structure was taken as a single CV. To further illustrate the generality of the approach, we consider the mean first passage time (MFPT) to the native node $B$ as an optimal RC. The optimization algorithm uses the MFPT optimization functional defined in Eq.~\ref{eq:mfpt_func} and variations as defined by Eq.~\ref{eq:var} with $\Delta_h=\{0, 1, 2, 4, 8, 16, 32, 64, 128, 256\}$, to compensate for the extremely low dimensionality of the input. Ten optimization runs were performed to ensure robustness, each initialized with a different random seed and run for 100,000 iterations.

\subsection{RC optimization without histories using a single CV}
\begin{figure}[h!]
	\centering
	\includegraphics[width=\linewidth]{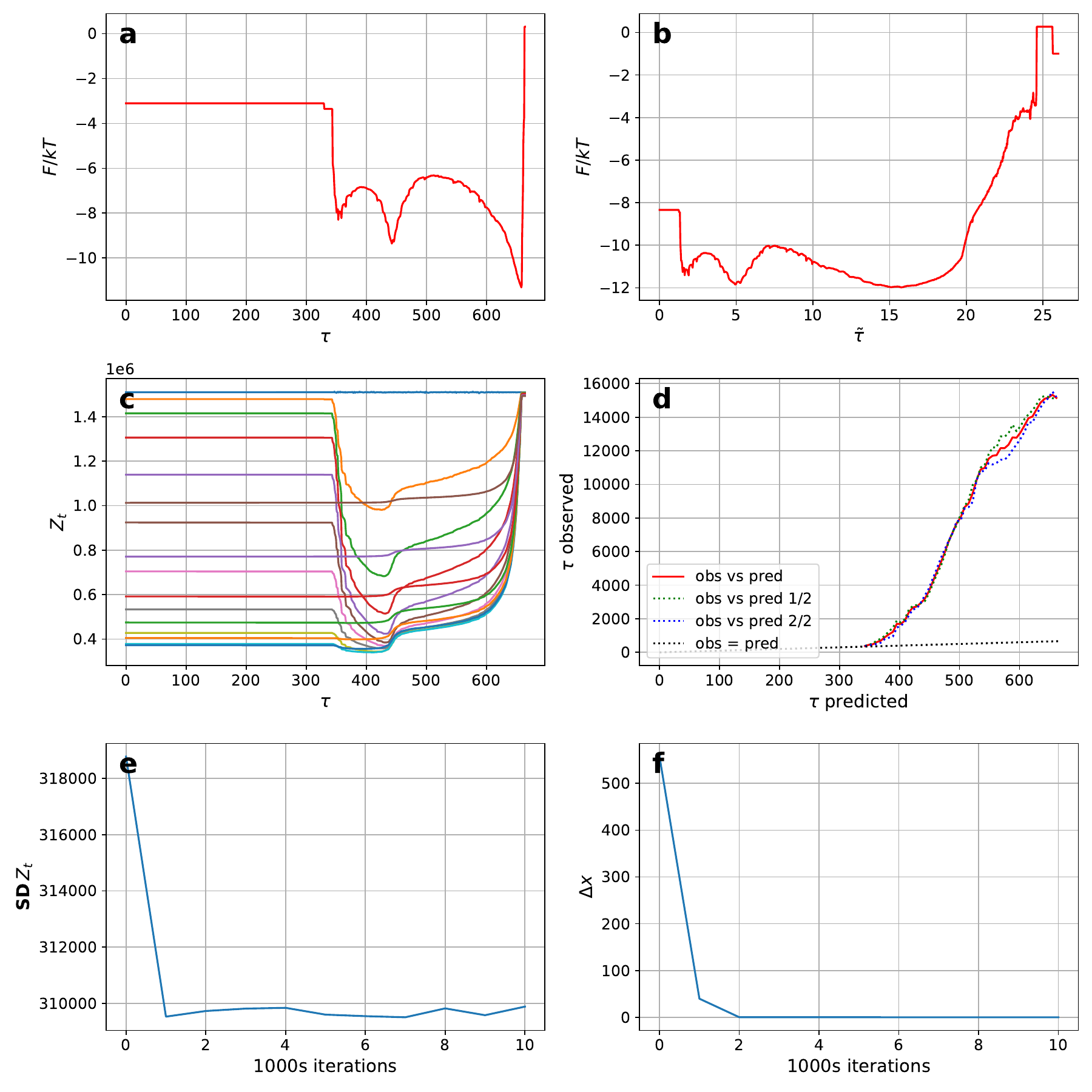}
	\caption{The MFPT RC optimization using a single variable - the rmsd time-series. Notation as in Fig. \ref{fig:4}.}
	\label{fig:S4}
\end{figure}

Supplementary Fig. \ref{fig:S4} shows the results of MFPT RC optimization without histories, using a single CV - the RMSD time-series. The results are significantly worse than those shown in Fig. \ref{fig:4}. In particular, the MFPT range in Supplementary Fig. \ref{fig:S4}a is nearly 30 times smaller than that shown in Fig. \ref{fig:4} and a similar discrepancy is observed for Supplementary Fig. \ref{fig:S3}b. The validation criterion (Supplementary Fig. \ref{fig:S4}c) shows significant deviations from a constant value, indicating the sub-optimality of the obtained RC and its poor approximation of the MFPT. This is further confirmed by the large discrepancy in the predicted vs. observed plot (Supplementary Fig. \ref{fig:S4}d). Supplementary Figs. \ref{fig:S4}e and \ref{fig:S4}f show that optimization converged in first 2000 steps.

\subsection{Langevin dynamics}
Langevin dynamics was simulated using the Euler-Maruyama method with the following parameters: mass $=1$, friction coefficient $\gamma=1$, Boltzmann constant $k_B=1$, temperature $T=1$, time step $\Delta t=0.01$, and a total of 1,000,000 steps.  The potential barrier was defined as $$U(x)=\exp[-(x-0.5)^2/0.01].$$ The resulting trajectory, which includes approximately 3,300 transitions between the boundaries, was analyzed using the optimization functional defined in Eq.~\ref{eq:func_final} and RC variations defined by Eq.~\ref{eq:varh}, with basis functions $f(y(t_{k-1}), y(t_k))$ and $f(y(t_k), r(t_k))$, using $x(t)$ as the sole CV. Optimization was performed for 10,000 iterations with $\gamma=0$. Once the time-series of optimal RC $q(t)$ was determined, the Langevin trajectory $x(t), v(t)$ was used to compute and visualize $q(x,v)$.

For backward committor, one considers time-reversed time-series of the CV $x_r(t)=x(T-t)$, the rest of the optimization parameters are the same.

\subsection{Conceptual ocean models}
The trajectory was analyzed using optimization functional defined in Eq.~\ref{eq:func_final} and RC variations followed Eq.~\ref{eq:varh}, with $\Delta_h=\{0,1\}$ and $A_i(t)$ as the input CVs.  The regularization parameter was set to $\gamma = 0.1$. Ten optimization runs were performed to ensure robustness, each initialized with a different random seed and run for 100,000 iterations.

\subsection{Longitudinal clinical dataset}
The clinical dataset was analyzed using (logarithm of) serum creatinine (sCr) time-series as a single input CV. 
Optimization was performed using the functional defined in Eq.~\ref{eq:func_final} and RC variations followed Eq.~\ref{eq:varh}, incorporating delayed inputs via the following basis functions
$f(y_m(t_l), r_m(t_l))$, $f(\ln (t_k-t_l), r_m(t_l))$, and $f(\ln (t_k-t_l), y_m(t_l))$, where $l=k-\Delta_h$.

Scenario I used $\Delta_h=\{0,1,2,3,4,5,6,7,8\}$, $\gamma=0.2$. Ten optimization runs were performed to ensure robustness, each initialized with a different random seed and run for 20,000 iterations. In this setting, the boundary conditions act as absorbing states (or traps), once a trajectory reaches a boundary, it terminates. To ensure that the validation criterion $Z_q$ remains constant for the optimal RC (the committor), it is necessary to extend trajectory segments to match the length of those that have not yet reached a boundary  \cite{banushkina2025data}. Since all trajectories in this scenario reach a boundary, they are effectively treated as infinitely long. 

Scenario II used $\Delta_h=\{0,1,2,3,...,8,9,10\}$, $\gamma=0.2$. Ten optimization runs were performed to ensure robustness, each initialized with a different random seed and run for 50,000 iterations. In this case the trajectories are not absorbed at the boundaries but continue evolving beyond them. Therefore no modification of transition path segment lengths is required.

\begin{figure}[ht!]
	\centering
	\includegraphics[width=\linewidth]{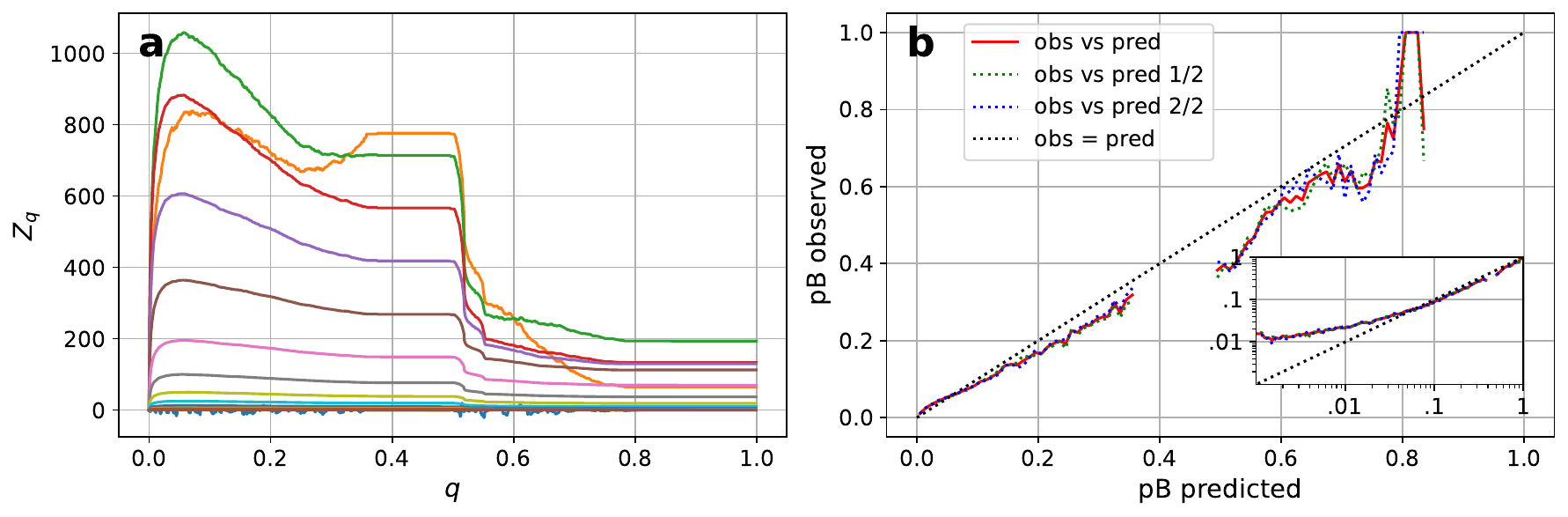}
	\caption{Committor optimization for AKI dynamics using a single sCr time-series without histories. Notation as in Fig. \ref{fig:7}.}
	\label{fig:S5}
\end{figure}

Supplementary Fig. \ref{fig:S5} shows the results obtained with optimization without histories, which are significantly worse than the results obtained using histories, as shown on Fig. \ref{fig:7}. In particular, the optimality criterion deviations from the constant value are more than an order of magnitude larger, and the predicted vs. observed plot exhibits significant disagreements, especially for low $q$ values. It is also discontinuous and fails to span the full range.
